\documentclass[a4paper,11pt]{article}
\pdfoutput=1
\usepackage[utf8]{inputenc}
\usepackage{graphicx}
\usepackage{amssymb}
\usepackage{amsthm}
\usepackage{multirow}
\usepackage{amsmath}
\usepackage{url,hyperref}
\usepackage{setspace,fullpage}
\usepackage{algorithm,algorithmic}
\usepackage[affil-it]{authblk}
\usepackage{lineno}

\DeclareMathOperator*{\argmax}{\arg\!\max}

\title{Better than Counting Seconds: Identifying Fallers among Healthy Elderly using Fusion of Accelerometer Features and Dual-Task Timed Up and Go\thanks{paper submitted to PLOS ONE, 2017}}
\author[1]{Moacir Ponti}
\author[2]{Patricia Bet}
\author[2]{Caroline L. Oliveira}
\author[2]{Paula C. Castro}
\date{}
\affil[1]{\small ICMC, Universidade de S\~ao Paulo -- S\~ao Carlos/SP, Brazil --- \url{ponti@usp.br}}
\affil[2]{\small DGero, Universidade Federal de S\~ao Carlos, S\~ao Carlos/SP, Brazil}
\begin{document}

\maketitle

\begin{abstract}
Devices and sensors for identification of fallers can be used to implement actions to prevent falls and to allow the elderly to live an independent life while reducing the long-term care costs. In this study we aimed to investigate the accuracy of Timed Up and Go test, for fallers’ identification, using fusion of features extracted from accelerometer data. Single and dual tasks TUG (manual and cognitive) were performed by a final sample (94\% power) of 36 community dwelling healthy older persons (18 fallers paired with 18 non-fallers) while they wear a single triaxial accelerometer at waist with sampling rate of 200Hz. The segmentation of the TUG different trials and its comparative analysis allows to better discriminate fallers from non-fallers, while conventional functional tests fail to do so. In addition, we show that the fusion of features improve the discrimination power, achieving AUC of 0.84 (Sensitivity$=$Specificity$=0.83$, 95\% CI $0.62$-$0.91$), and demonstrating the clinical relevance of the 
study. 
We concluded that features extracted from segmented TUG trials acquired with dual tasks has potential to improve performance when identifying fallers via accelerometer sensors, which can improve TUG accuracy for clinical and epidemiological applications.
\end{abstract}



\section{Introduction}

   Several consequences arise with increasing elderly population, among them the intensified possibility of occurrence of falls~\cite{Almeida2012}. The World Health Organisation defines fall as ``come to inadvertently get in the soil or in other lower level, excluding intentional position changes to lean on furniture, walls or other objects''~\cite{World2008}. More than a third of older persons fall at least one time per year~\cite{World2008}. Half of those who fell once are likely to experience other falls in the following months, with physical and functional consequences, such as pain, bone fractures, mobility disability, amputations, institutionalisation and cost increases with health care~\cite{Palvanen2014}. These factors put falls as a public-health problem of great importance~\cite{Delbaere2010,Cigolle2015,Rosa2015}.
   
   Devices that help identify fallers can be used to develop programs and implement actions to prevent these falls and to allow the elderly to live an independent life while reducing the long-term care costs. Signals obtained by small wearable sensors are widely studied for this purpose. Because those are designed to be comfortable to use, those signals can be acquired in high sampling rates and even for long periods, making them a suitable choice to assess ageing in several applications, including fall risk and faller identification for ageing studies~\cite{Rashidi2013,Weiss2013,Ponti2015learning,salarian2010itug,zampieri2010instrumented,zampieri2011assessing}. Albeit these facts, the accuracy is key to enable clinical, public health and epidemiological research uses of the signals.
   
   Because of the strong appeal of the application, several papers report measures and features that can be extracted from inertial sensors --- such as gyroscopes and accelerometers --- and also laser, cameras and devices with multiple sensors~\cite{Sejdic2014,Yuwono2014,Millecamps2015}. In this context the accelerometers are specially valuable due to be small, cheap, easy to wear and with low power consumption when compared to more complex devices. Recently, regular triaxial accelerometers were found to be reliable when compared with legacy equipments~\cite{Byun2016}. Moreover they do not depend on environmental monitoring, minimising problems with privacy, usually caused by cameras and home based sensors. In this study we are interested on the analysis of accelerometer data obtained from a single triaxial sensor to identify the fallers. 
  
   Among the gait related studies using accelerometer sensors, Pogorelc et al.~\cite{Pogorelc2010,Pogorelc2013} focus on detecting health conditions in older persons such as Parkinson's disease and others, from human gait. Capela et al.~\cite{Capela2015} address activity identification on the elderly. However, while the detection of fall events was extensively studied~\cite{Bagala2012, Amin2015}, the identification of fallers and non-fallers via accelerometer data, in particular with a single sensor, remains an open problem~\cite{Ejupi2014}. In community-dwelling elderly this poses a bigger challenge, since conventional functional tests such as Timed Up and Go (TUG) tests have limited ability to predict falls~\cite{Barry2014}.
   
   In a survey about fall risk, the authors pointed out the important role of the sitting and standing movements in the continuous monitoring of functional mobility~\cite{Ejupi2014}. Indeed, recent studies about fall and technology often apply Timed Up and Go (TUG) tests with one or more accelerometer sensors~\cite{Doi2013,Doheny2013,Liu2011,Gietzelt2009} and also gyroscopes~\cite{Greene2010} in order to investigate gait behaviour and falls, often in hospitalised or disabled participants. Furthermore, Salarian et al.~\cite{salarian2010itug} developed an iTUG test using from five to seven accelerometer sensors; which had good psychometric properties at a pilot study for Parkinson's patients, besides being fit for home assessment~\cite{zampieri2011assessing} at early stages of the disease. Main features that demonstrated association with UPDRS (Unified Parkinson's disease rating scale), extracted from iTUG are step counting, seconds, peak arm velocity, cadence, stride and turning and among the subcomponents of 
iTUG, gait, turning, and turn-to-sit were the most reliable.
   
   Regarding the feature extraction from accelerometer signals, previous studies explored time, frequency and time/frequency features. In~\cite{Weiss2011}, the use of frequency features, in particular the amplitude of the frequency peaks, showed different means between the groups faller and non-faller. In a study using the use of accelerometer sensors during one week, the dominant frequencies were associated with the fall history~\cite{vanSchooten2015}. One of the most interesting related studies uses of a pair of accelerometers to collect data freely in daily-life activities for 3 days in order to extract features related to fall risk~\cite{Weiss2013}. However, it is based on the detection of steps, which we want to avoid in our study, using instead features that can be extracted directly from the acceleration data. 
   
   Human gait on healthy adults can be well represented by frequencies up to 15Hz for walking, running and jumping~\cite{Bhattacharya1980}. There is still little evidence about the adequate bandwidth for acquiring accelerometer signals from older adults in activities such as walking, sitting, standing and free movement~\cite{Mannini2013,Ejupi2014}. Those were already studied for healthy adults but not yet established for elderly gait and fall patterns~\cite{Murphy2009, Sekine2000}.   

   In this paper, we present results of feature extraction on accelerometer data with the aim of identifying fallers among a group of healthy community-dwelling older than 60 years adults  (socially engaged, robust, active, non-obese, with preserved cognition). The fact that the sample is composed by healthy older persons makes it more difficult to identify fallers by using regular functional and screening tests such as the TUG time trial~\cite{Alexandre2012}. We study features extracted from accelerometer data collected during three consecutive TUG tests, in particular frequency features, and analyse how those features can be used to discriminate between faller and non-fallers. In this study we use only one accelerometer, and three TUGs (single and dual-task manual/cognitive). Therefore, we extract features considering the whole signal --- composed by the acceleration data of 3 consecutive TUGs --- but also considering individually each TUG trial.

   Our main contributions are: 
   \begin{enumerate}
   \item the collection of an open dataset for faller identification, not currently available in the literature, allowing reproducibility and comparison with future studies;
   \item the study of faller identification problem using: TUG single task, dual-task manual TUG (TUG-M) and dual-task cognitive TUG (TUG-C), not yet explored in the literature in the context of accelerometer-based faller identification;
   \item the description of features based on TUGs and differences between TUGs using a single accelerometer sensor, as well as feature fusion methods, resulting in variables that are able to discriminate fallers from non-fallers while conventional functional tests cannot.
   \end{enumerate}
   
   As far as we know there is no previous study including the items 2-3 described above, i.e. that compares gait features using TUG variations for the purpose of elderly faller's identification using a single accelerometer sensor, independently of step detection and investigating adequate sampling rates. It is important to emphasise that our sample is not composed of participants with diseases that can impair gait, such as Parkinson's Disease in~\cite{zampieri2011assessing} and related literature, and that we use as input a single accelerometer sensor.
   Also, studies on fall risk and faller identification often do not release the dataset for reproducibility in those studies, and their sample are seldom stratified with respect to gender and age, often not providing information on functional mobility of the subjects~\cite{Ejupi2014}. We believe that making this dataset available will allow for a faster advance of the field. Finally, as we show in the results, the identification of fallers among a sample of health elderly is challenging and conventional functional tests fail to provide a threshold for screening.


\section{Method}
\label{s.method}

The overall schemes are shown in Fig~\ref{fig:flowchart1} and Fig~\ref{fig:flowchart2}, respectively describing the data collection and processing.  

\subsection*{Study Subjects}
\label{ss.datasampling}

This section describes the subjects sampling, not to be confused with the accelerometer sampling rate (for which we give details later in the text). 

\begin{figure}
\centering
     \includegraphics[width=.8\linewidth]{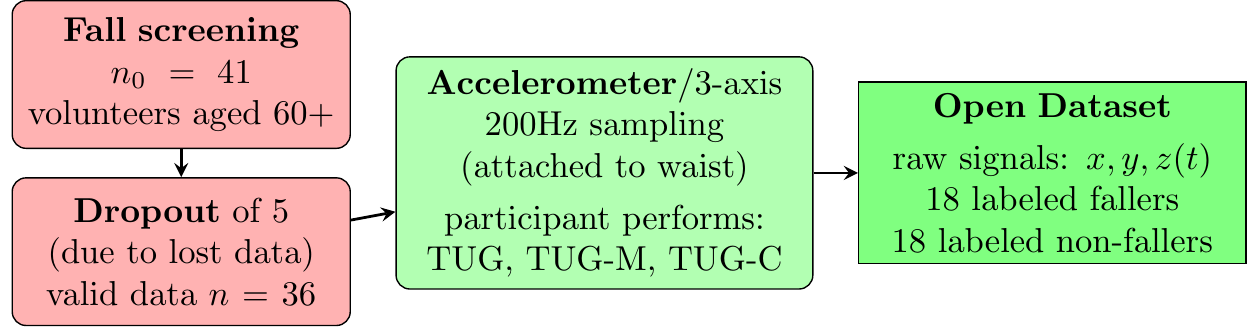}
\caption{Flowchart of the dataset collection process.}
\label{fig:flowchart1}
\end{figure}

This study enrolled 41 community-dwelling elderly divided as: 19 fallers and 22 non-fallers volunteer participants’ residents in S\~ao Carlos/SP, Brazil, in 2015. The history of falls was determined with a single question: ``have you fallen within the past year?''. Data from five participants were lost due to acquisition issues; thus, this study final sample (94\% power and 5\% error) is 18 fallers and 18 non-fallers, as depicted in Fig~\ref{fig:flowchart1}.

The study protocol (482.306/2013) and informed consent form received ethics approval from the UFSCar Ethic Committee on Human Experimentation. The project was advertised at the University of the Third Age groups, for a population of 468 elderly and the volunteers received a written informed consent concerning conduct of the survey. Participation was voluntary and it was explained that the volunteer could leave the study whenever he/she wanted without suffering any loss or consequence. In the open dataset, only the raw accelerometer signal, the gender and the label (faller or non-faller) are available, while the remaining variables are not available in order to assure complete anonymity of participants. All the participants were refer to free Physical Therapy and Gerontology intervention after this study. 

The eligibility criteria for the study included participants being 60 years and older; possessing the ability to stand up from a chair with arms without other person's help and to walk independently, without aid device. 
The exclusion criteria were: amputation and/or use of lower limb prosthesis or other device that modifies the gait pattern; neurological or muscular disease, any condition listed it Charlson Comorbidity Index \cite{Martins2008} and presence of any important risk factor that compromises safety, perceived by the evaluator, such as blood pressure lower than 90/60 or higher than 140/100 mmHg or angina. All the participants were eutrophic and none of them was obese \cite{Vasconcelos2010}.

The Table \ref{tab:sampledescription} summarizes the demographic and functional mobility characteristics of this study participants: MMSE (mini-mental state examination) \cite{Lourenco2008}, FES (Falls Efficacy Scale---Brazil) \cite{Camargos2010} , the three TUGs \cite{Silsupadol2006}: (i) single task, (ii) TUG-M (dual task manual), (iii) TUG-C (dual task cognitive). The TUG results are shown in seconds. We also compared the differences in seconds between the different TUG tests, computed the average between the TUG seconds for each participant, and counted the number of steps during each test. The groups were compared using the $t$-test for all variables, except for the Gender, for which a Fisher's exact test was used.

\begin{table}[hpbt]
\caption{Sample description, including the results of MMSE, FES-1-Brazil, TUG, TUG-M, TUG-C and $p$-values for statistical tests comparing the groups: non-faller and faller.}
\label{tab:sampledescription}

\centering
\begin{tabular}{c c | c c c}
 \hline
        &                 & non-faller & faller & $p$-value\\ 
        \hline  \hline
n       &                 & 18              & 18            & \\
Age     & $\mu\pm\sigma$  & $70.94\pm6.69$  & $75.25\pm8.20$& 0.102$^a$\\
\multirow{ 2}{*}{Gender}  & Female          & 10 (56\%)       & 15 (83\%)     & \multirow{ 2}{*}{0.146$^b$}\\
        & Male            & 8 (44\%)        & 3 (17\%)      & \\
MMSE    & $\mu\pm\sigma$  & $26.46\pm4.35$  & $23.75\pm3.93$& 0.117$^a$\\
FES     & $\mu\pm\sigma$  & $24.62\pm7.74$  & $21.00\pm6.55$& 0.222$^a$\\
TUG (s)  & $\mu\pm\sigma$ & $9.026\pm2.376$ & $10.395\pm2.713$ & 0.094$^a$\\
TUG-M (s)& $\mu\pm\sigma$ & $9.790\pm3.016$ & $10.974\pm2.713$ & 0.232$^a$\\
TUG-C (s)& $\mu\pm\sigma$ & $13.806\pm5.962$& $17.016\pm6.250$ & 0.130$^a$\\
\hline
TUG avg (s) & $\mu\pm\sigma$ & $10.870\pm3.480$ & $12.790\pm3.261$ & 0.067$^a$\\
TUG-M - TUG (s)  & $\mu\pm\sigma$ & $1.028\pm1.055$ & $0.972\pm0.804$ & 0.863$^a$\\
TUG-C - TUG (s)& $\mu\pm\sigma$ & $4.201\pm4.361$ & $6.123\pm5.281$ & 0.248$^a$\\
TUG-M - TUG-C (s)& $\mu\pm\sigma$ & $4.872\pm4.307$ & $6.677\pm5.355$& 0.279$^a$\\
\hline
steps TUG   & $\mu\pm\sigma$ & $13.954\pm3.261$ & $16.111\pm4.114$& 0.090$^a$\\
steps TUG-M & $\mu\pm\sigma$ & $14.500\pm2.727$& $15.500\pm3.976$ & 0.430$^a$\\
steps TUG-C & $\mu\pm\sigma$ & $17.012\pm4.958$ & $18.556\pm6.608$& 0.387$^a$\\
 \hline
 \multicolumn{5}{l}{{\footnotesize $a$ --- $p$-values for $t$-test;  $b$ --- $p$-value for Fisher's exact test.}}
\end{tabular}
\end{table}

The groups Faller and Non-faller were paired in gender and age to allow comparison. Therefore, the groups do not have significant differences considering the demographic characteristics. Even so, the fallers display a expected slightly older average age since fall prevalence increases with age~\cite{Cigolle2015}. The groups are also, as expected, more feminine, probably because, as they age, women are more likely to become fallers and to experience negative outcomes from a fall episode than men~\cite{Rosa2015}.

Both groups are similar also regarding the functional mobility variables. According to the Falls Efficacy Scale---Brazil (FES-I-Brazil) \cite{Camargos2010}, participants from both groups feel little concern with the possibility of falling when carrying out functional activities. Moreover, while it is expected that the fallers conduct the TUG single task in more than 12.47 seconds~\cite{Alexandre2012}, this particular group of Fallers equates to non-fallers for TUG execution time. Even for TUG Manual and TUG Cognitive times, both groups can be considered similar. 
Considering only the functional mobility tests would {\bf not} be possible to discriminate between fallers and non-fallers. This is probably because the participants are involved in regular physical and cognitive activities and can be considered in successful ageing~\cite{Rowe2015}. In addition, the control variables addressed by the exclusion criteria are related to increases in fall~\cite{Delbaere2010}~\cite{vanSchooten2015}. Therefore, other methods are needed to identify fallers in this scenario.

\subsection*{Timed Up and Go tests (TUG)}
\label{ss.tug}
The Timed Up and Go Test (TUG) is widely used in both clinical and epidemiological studies; since the time spent to complete the test is often correlated to functional mobility and associated with a past history of falls~\cite{Beauchet2011}. The TUG is also used to assess the risk of falls and to select interventions for older individuals according to the updated guidelines of the American and British Geriatric Societies for the Prevention of Falls~\cite{Kenny2011}. For Brazilian older adults, the 12.47 seconds cut-off point is adopted as a predictive value for fall \cite{Alexandre2012}.

While both single and dual tasks TUG have shown good psychometric properties, dual task TUG with manual or cognitive components have potential to provide information even considering the limiting compensation for healthy older persons~\cite{Hofheinz2010}, probably because cognitive and motor reserve can influence the gait efficiency~\cite{Hausdorff2008}.  Because single task TUG tests have shown limited ability to predict falls among community-dwelling elderly~\cite{Barry2014}, and since dual task TUG presents good psychometric results, highly standardised administration procedures and it is easily applicable considering its low cost, time and space required, many studies have successfully investigate identification of clinical important conditions, such as frailty and disabilities risk in patients with chronic conditions using manual or cognitive dual task TUG~\cite{Tang2015}~\cite{Weiss2013using}~\cite{Vance2015}.

\begin{figure}
\centering
     \includegraphics[width=.9\linewidth]{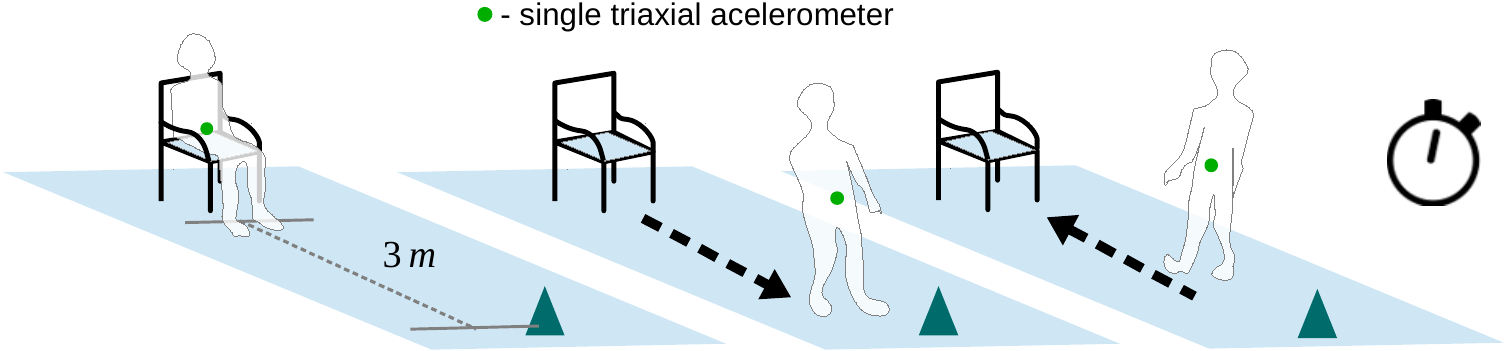} 
\caption{Timed Up and Go test (TUG) illustration}
\label{fig:TUGexample}
\end{figure}

In this study, three variants of TUGs were conducted by two trained gerontologists researchers. The regular one we refer as single task or only TUG, requiring a participant to stand up from a chair, walk 3 meters, turn, walk back, and sit down, while the time taken to perform the task is recorded using a stopwatch. This procedure is illustrated in Fig~\ref{fig:TUGexample}. In addition the TUG dual-task is adopted in two different approaches: TUG Manual (TUG-M), following the same procedure as the regular TUG, but carrying a cup filled with water; and TUG Cognitive (TUG-C), in which participants are asked to respond to continuous simple subtraction questions while performing the TUG test \cite{Silsupadol2006}~\cite{Shumway2000}.

In order to analyse the gait patterns during the TUG tests, instead of using just the seconds or the number of steps taken to perform the tests, a third independent researcher (blind) extracted features based on information theory such as entropy in order to measure complexity, and frequency features that tries to analyse periodicity, speed and stability of the acceleration by looking at the peak frequencies and harmonics~\cite{Sejdic2014,Orovic2011}. 

\subsection*{Signal acquisition and pre-processing}
\label{ss.dataacquisition}

\begin{figure*}[hbt]
\centering
     \includegraphics[width=1\linewidth]{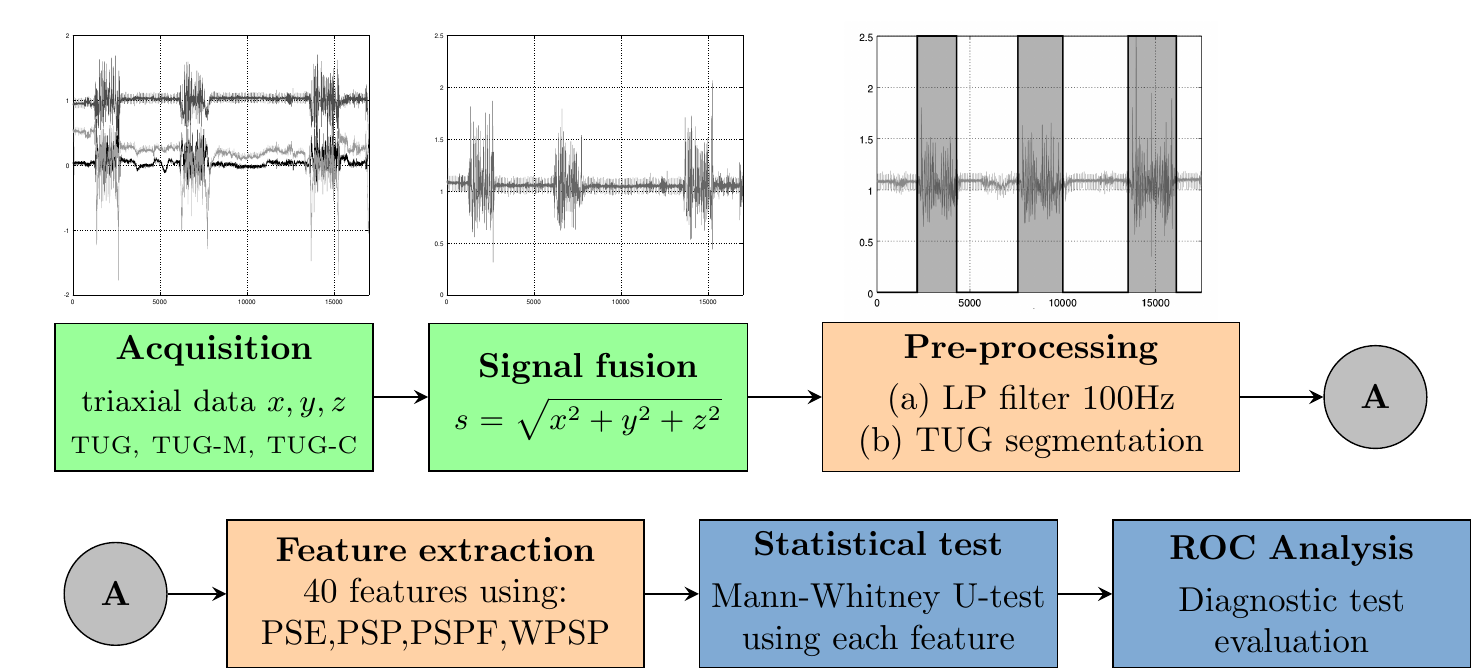}
\caption{Overall picture of the methodology: the triaxial signal is collected and a fusion is performed to keep the methods orientation-independent; then the signal is filtered using Butterworth method, and each TUG is segmented, and 32 features are extracted; the signal is analysed with respect to the sampling rate, the extracted features are studied using a statistical test, a feature ranking and a classification experiment.}
\label{fig:flowchart2}
\end{figure*}

A single triaxial accelerometer sensor (Analog Devices ADXL362) was used to acquire the signal using a datalogger set at a sampling rate of 200Hz. Each participant was ask to wear the sensor using an elastic belt around his/her waist (in front of the mass centre). In Fig~\ref{fig:flowchart2} the overall pipeline is shown: the two first steps (green boxes) are related to the data acquisition; the following two (orange boxes) to the signal pre-processing and feature extraction, and highlighted in darker blue shade are the results obtained using the statistical tests and ROC curve analysis. The details of each step are given in the following sections.

\paragraph{Signal fusion} all feature extraction and measurements are obtained from $s(t)$, which is computed as:
\begin{equation*}
s(t) = \sqrt{x(t)^2 + y(t)^2 + z(t)^2},
\end{equation*}
where $x(t)$,  $y(t)$ and $z(t)$ are the accelerometer data acquired from the axis $x$, $y$ and $z$, respectively. Because we do not assume a fixed position of the sensor, the squared sum of each axis allows comparing different outputs regardless the sensor orientation.

Note that the discrimination between fallers and non-fallers is not trivial by looking only at the signals $s(t)$. Therefore, pre-processing and frequency analysis are needed to extract discriminative features, as we will describe in the following sections.

\paragraph{Low-pass filter} in studies with accelerometers it is common to apply a Butterworth low-pass filter in order to suppress noise using bandwidth values such as: 50Hz~\cite{Greene2010}, 20Hz~\cite{Doi2013} and 5Hz~\cite{Doheny2013}. Human gait for healthy adults was found to be well characterised by frequencies up to 15Hz for walking, running and jumping~\cite{Bhattacharya1980,Mannini2013}, but there is not enough evidence on the elderly specially when doing tasks such as sitting and standing (as in the TUG test). 
 
By the Nyquist-Shannon sampling theorem, if a signal is band-limited by a frequency $B$, a sampling rate of $2B$ samples per second is needed in order to perfectly reconstruct the signal~\cite{Nyquist1928}. Since the use of an arbitrary low pass filter may hamper the analysis, we use a Butterworth filter of 100Hz as an anti-aliasing filter for frequencies higher than $(200/2)$Hz, because it has the least attenuation over the desired frequency range~\cite{Baker1999}.

\paragraph{Frequency analysis} is a widely used tool to extract information from signals that are not clear when looking at the time domain. We use the Fast Fourier Transform in order to analyse the frequency characteristics of the acquired signals, with respect to the different TUG tests.

\subsection*{Signal segmentation}
Because one of the contributions of this study is to acquire data from 3 TUGs, we need to separate those trials by segmenting the signal before the feature extraction step. Although it is possible to consider other segmentation methods, we present a simple algorithm to segment the consecutive TUG trials.

An example of segmentation is shown in the third step of Fig~\ref{fig:flowchart2}, in which the regions of the signal outside the grey regions will be ignored in the feature extraction step. 

The segmentation is described in Algorithm~\ref{alg:ss}, which basically computes the mean to be subtracted from the input signal (lines 1-2), applying a rolling median filter in order to reduce variance within each TUG (line 3) and then, for each half second, sums the values contained in the processed signal (lines 4-5), then it thresholds the data by using this sum (lines 6-9). Finally, each segment is labelled and this result is returned (lines 12-13). This algorithm succeeded to segment all but two signals, for which the segmentation had to be corrected manually.

\begin{algorithm}
	\caption{Signal segmentation}
	\label{alg:ss}
	\begin{algorithmic}[1]	
	\REQUIRE $k$ to compute the rolling median filter 
	\STATE $\bar s \leftarrow \operatorname{mean}(s)$
	\STATE $s \leftarrow s - \bar s$
	\STATE $u \leftarrow \operatorname{rollingMedianFilter}(s,k)$
	\FOR {each half second $h$ in $u$}
	    \STATE $g(h) \leftarrow\sum_h{u(h)}$
		\IF {$g(k) > \bar s$}
			\STATE $l(k) = 1$
		\ELSE
    	    \STATE $l(k) = 0$
        \ENDIF
	\ENDFOR
    \STATE assign distinct labels to each segment in $l$
	\RETURN {$l$}
	\end{algorithmic}
\end{algorithm}

\subsection*{Features}
\label{ss.features}

Let $n$ be the size of a given signal $s$, and its power spectrum $S(\omega) = |\mathcal{F}(s)|^2$. Four frequency domain features are used as follows. 
\begin{enumerate}
    \item Power Spectral Entropy (PSE): the entropy of the power spectrum of the signal. It represents a measure of energy compaction in transform coding~\cite{Gibson1994}, in our case how much acceleration energy the signal contains.
    \begin{equation}
    f_1 = -\sum_{\omega} S(\omega) \cdot \log(S(\omega)+\epsilon),
    \end{equation}
    where $\epsilon = 0.001$ to avoid $\log(0)$;
        
    \item Power Spectrum Peak Frequency (PSPF): computed by finding the frequency related to the higher value of $S$. This feature represents the first harmonics of the gait, which is related to the overall movement speed:
    \begin{align}
    f_{2,1} &= \argmax_{\omega} S(\omega);\\
    f_{2,2} &= \argmax_{\omega- \left\lbrace f_{2,1}\right\rbrace} S(\omega);\\
    f_{2,3} &= \argmax_{\omega - \left\lbrace f_{2,1},f_{2,2}\right\rbrace} S(\omega) .
    \end{align}

    \item Power Spectrum Peak (PSP): computed by finding the three highest values of $S$. This third feature represents the amplitudes of the fundamental frequencies of the gait:
    \begin{align}
    f_{3,1} &= S(f_{2,1});\\
    f_{3,2} &= S(f_{2,2});\\
    f_{3,3} &= S(f_{2,3}).
    \end{align}
    
    \item Weighted Power Spectrum Peak (WPSP): computed using the PSP values weighed by the PSPF values
    \begin{align}
    f_{4,1} &= f_{2,1} \cdot f_{3,1};\\
    f_{4,2} &= f_{2,2} \cdot f_{3,2};\\
    f_{4,3} &= f_{2,3} \cdot f_{3,3}.\\
    \end{align}    
    
\end{enumerate}

Each feature described above is extracted from the following signals:
\begin{enumerate}
    \item[s)] The whole signal (containing the three TUG trials);
    \item[t)] The first TUG trial (TUG);
    \item[m)] The second TUG trial (TUG-M);
    \item[c)] The third TUG trial (TUG-C).
\end{enumerate}

Thus, we have a total of 40 features, i.e. 10 features extracted from each one of the 4 signals, that compose the final feature vector

\begin{align*}
\mathbf{x}_i = \lbrace
  & f_{1}^{s}; f_{2,(1:3)}^{s}; f_{3,(1:3)}^{s}; f_{4,(1:3)}^{s};\\
  & f_{1}^{t}; f_{2,(1:3)}^{t}; f_{3,(1:3)}^{t}; f_{4,(1:3)}^{t};\\ 
  & f_{1}^{m}; f_{2,(1:3)}^{m}; f_{3,(1:3)}^{m}; f_{4,(1:3)}^{m};\\ 
  & f_{1}^{c}; f_{2,(1:3)}^{c}; f_{3,(1:3)}^{c}; f_{4,(1:3)}^{c};
 \rbrace
\end{align*}

\paragraph{Distance-based features}
Because we are also interested in understanding how the TUGs with additional tasks are different from the regular one, we also computed  the Euclidean distance $d_j(.,.)$ for each feature $j$ related to the full signal (s) and the first TUG (t), to the two other TUGs --- (m) manual task, (c) cognitive task:

 \begin{align}
     d_j(s,t) &= \sqrt{\left[ f_{j}^s - f_{j}^t\right]^2},\\
     d_j(s,m) &= \sqrt{\left[ f_{j}^s - f_{j}^m\right]^2},\\
     d_j(s,c) &= \sqrt{\left[ f_{j}^s - f_{j}^c\right]^2},\\
     d_j(t,m) &= \sqrt{\left[ f_{j}^t - f_{j}^m\right]^2},\\
     d_j(t,c) &= \sqrt{\left[ f_{j}^t - f_{j}^c\right]^2},
\end{align}
for each feature $j = 1\cdots10$. We then test the distances in order to see if they showed significant different between the groups.

\paragraph{Feature fusion}
We also performed fusion of relevant features by using the normalised average. This practice is common in pattern recognition systems in order to take advantage of the complementarity of different variables. The average method is known for producing predictions with reduced variance, which can potentially improve the results~\cite{Ponti2011a}. This combination is often called early fusion and in this paper it is performed by first normalising each variable to a $[0,1]$ range and then averaging all values, obtaining a single combined variable.

\subsection*{Feature analysis}
\label{s.featureanalysis}
In order to look for the best features we use two approaches.

{\bf Statistical test} --- Mann-Whitney U-test is carried out in each feature in order to compare the faller and non-faller groups. The U test was chosen because we do not have information about the distribution of the variables. 

We individually evaluate each variable/feature to look for the ones that are able to discriminate fallers from non-fallers, i.e. rejects the null hypothesis of equal means. We also performed fusion of relevant features by using the normalised average. This practice is common in pattern recognition systems in order to take advantage of the complementarity of different variables. The combination using the average is often called early fusion~\cite{Ponti2011a}.

{\bf ROC analysis} --- In order to compare how the variables extracted from the different signals are able to provide discrimination between the groups, we performed ROC (Receiver Operation Characteristic) analysis~\cite{Fawcett2006}. A ROC curve shows the relationship between the True Positive Rate (TPR) and the False Positive Rate (FPR). Those values are related to the Sensitivity (TPR) and the Specificity (1-FPR).

In addition the following values are analysed: the Area Under the ROC curve (AUC), which can be interpreted as the probability that the classifier will rank a randomly chosen positive instance higher than a randomly chosen negative instance; and the f1-Score, which is the harmonic mean between the precision and the  sensitivity. We perform also a Sensitivity vs Specificity analysis, in order to find the optimum probability threshold for the variable.

\subsection*{Reproducibility}
The data and the code used to produce the results are available at \url{https://github.com/maponti/Gait-Analysis-for-Faller-Identification}. The dataset is anonymised and contains the extracted features. The code will allow researchers to extract the features in their own data, as well as reproduce our analyses.

\section{Results and Discussion}
\label{s.results}

\subsection*{Statistical test in individual features}
\label{ss.utestresults}
\begin{figure}[hbt]
\centering
\setlength\tabcolsep{1.5pt}
    \begin{tabular}{cc}
    \includegraphics[width=0.48\linewidth]{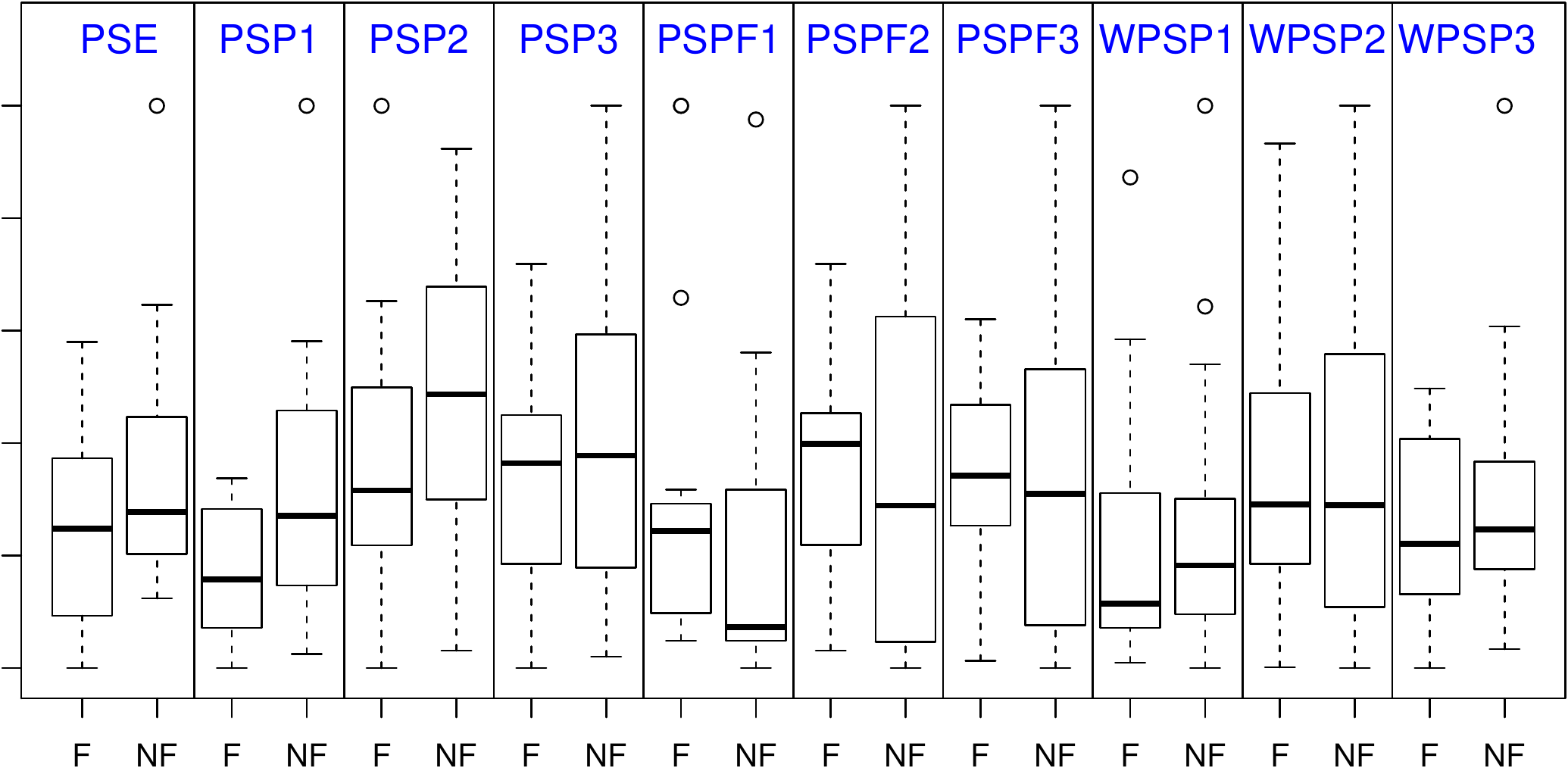} &
    \includegraphics[width=0.48\linewidth]{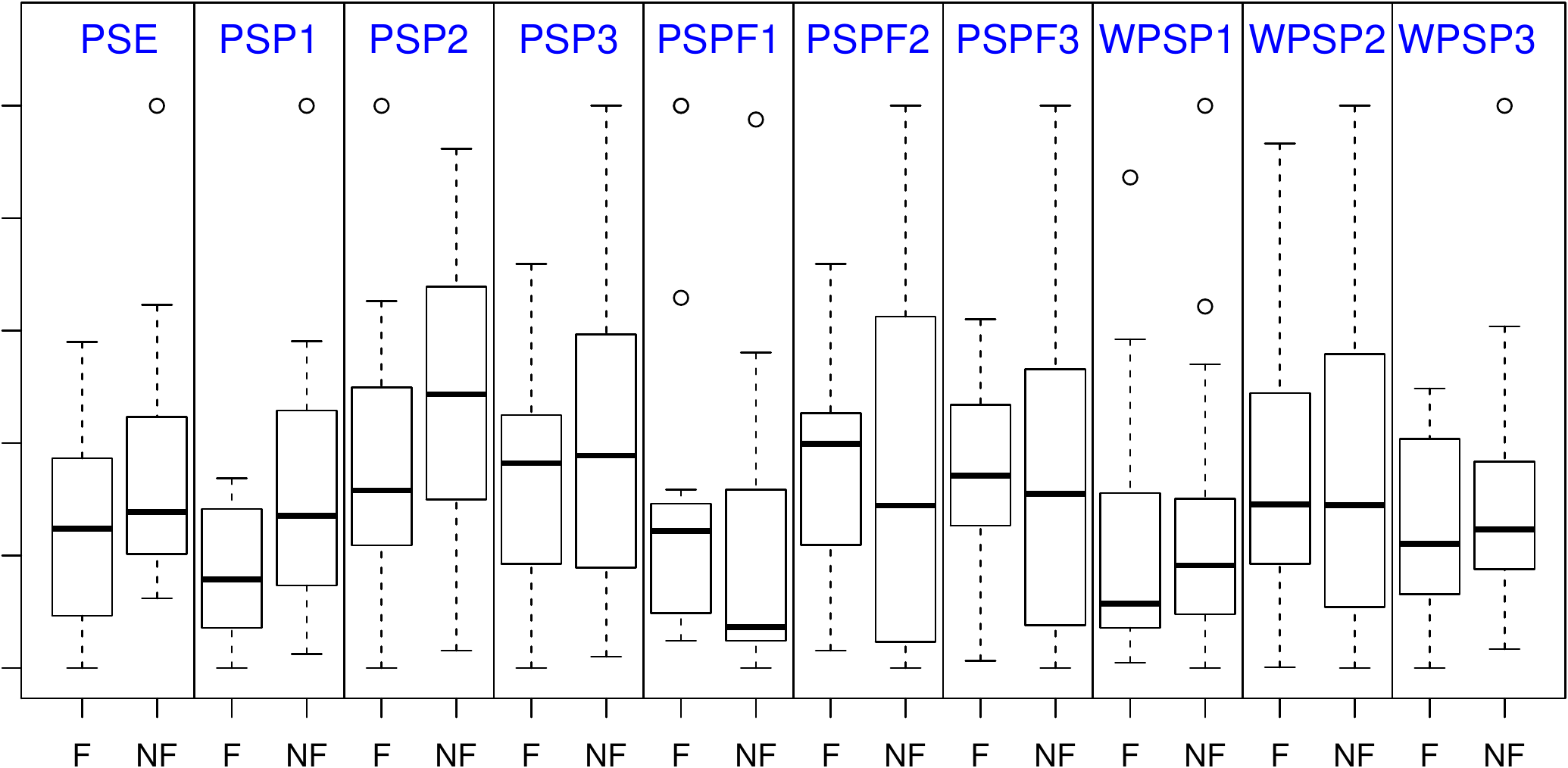} \\
    Signal (TUG+TUG-M+TUG-C) &  TUG \\[12pt]
    \includegraphics[width=0.48\linewidth]{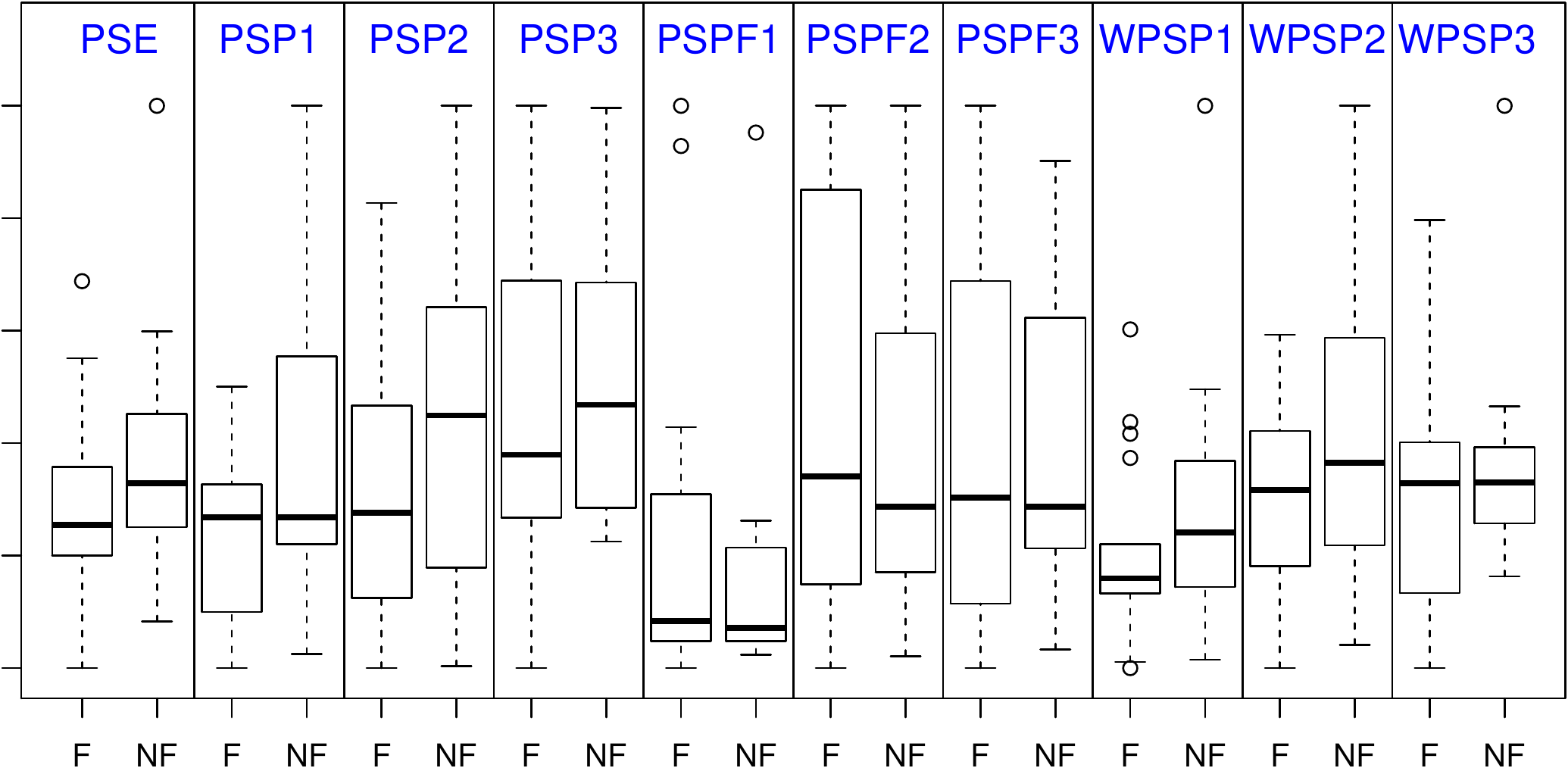} &
    \includegraphics[width=0.48\linewidth]{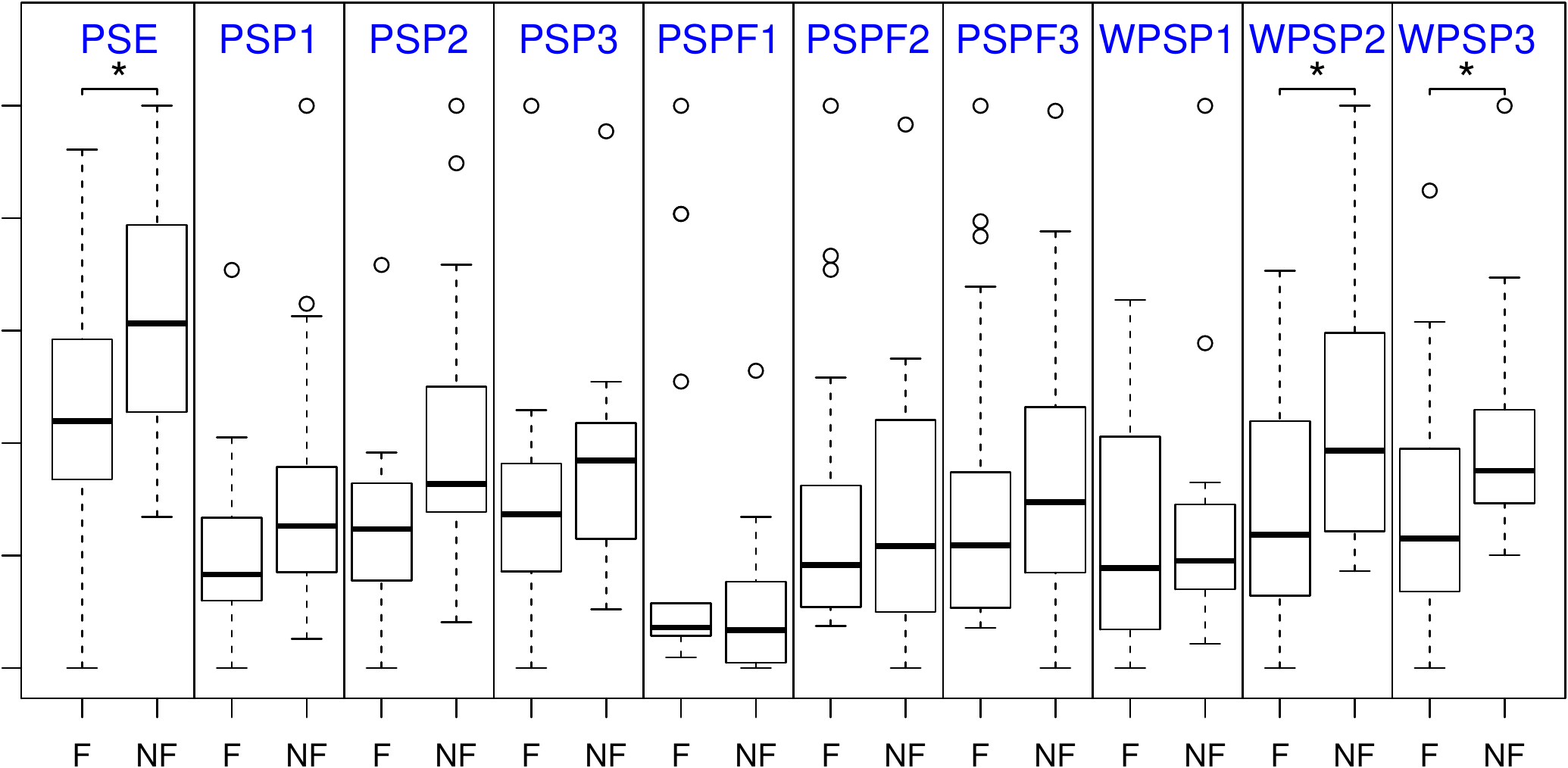} \\
    TUG-M &  TUG-C
    \end{tabular}
\caption{Frequency feature comparison when extracted from the complete signal, and individually from the TUG, TUG-M and TUG-C. OBS: values are normalised to the same numeric range for visualization; * indicates statistical significance for $p \leq 0.05$.}
\label{fig:featsboxplot}
\end{figure}

By using the statistical test, we can assess the capability of a given feature (variable) to reject the null hypothesis of equal means between the faller and non-faller groups. We are interested in features that are able to better discriminate between groups, in contrast with the regular functional tests which had failed to show significant differences. In Figure~\ref{fig:featsboxplot} we show the boxplots of each group for each feature (in order to visualise all in the same plot, the values are normalised to the same numeric range). It is possible to observe that the frequency features seems to be informative. However, the statistical test showed differences only for the features extracted from the TUG-C, namely the PSE and the WPSP2 and WPSP3.

The mean of each group ($\mu_{+}$ fallers, $\mu_{-}$ non-fallers) and the $p$-value for the statistical significant features (for $p \leq 0.05$) are shown in Table~\ref{tab:utestres}. Note that we also included the results for the normalised average variables obtained by combining both the relevant frequency features, and the relevant distance-based features.

As mentioned before, the only individual variables showing significance were those computed using only the TUG-C. In addition, we could find other variables by using the distances between features; the significant ones were: PSE and PSP differences between the whole signal (s) and the TUG-C (c), PSPF difference between TUG (t) and TUG-M (m), and finally the WPSP difference between TUG-M (m) and TUG-C (c). Our feature fusion method using the normalised average was also shown to be effective --- the combined variable also shows significance when comparing the groups.

\begin{table}[hpbt]
\caption{Features that produced statistical difference when comparing the means of the faller and non-faller groups}
\label{tab:utestres}
\centering
\begin{tabular}{c c | r r r}
 \hline
      &  & $\mu_{+}$ (fallers) & $\mu_{-}$ (non-fallers) & $p$-value\\ 
 \hline
 \hline
1 & PSE-$c$        &  $9.52\pm2.50$     & $11.95\pm2.90$      & 0.014 \\
2 & WPSP-$c$,2     &  $1.00\pm0.70$     & $1.73\pm0.90$      & 0.022 \\
3 & WPSP-$c$,3     &  $1.61\pm0.60$     & $2.30\pm0.80$      & 0.009 \\
\hline
\multicolumn{2}{c|}{Features fusion : avg(1,2,3)}
                   & $0.324\pm0.154$ & $0.500\pm0.161$  & 0.001 \\[2pt]
\hline
\hline
4 & $d_{\text{PSE}}(s,c)$ &  $6.689\pm2.48$     & $8.497\pm2.09$  & 0.029 \\
5 & $d_{\text{PSP}}(s,c)$ &  $0.024\pm0.01$     & $0.039\pm0.01$  & 0.014\\
6 & $d_{\text{PSPF}}(t,m)$&  $17.77\pm6.13$     & $21.38\pm4.28$  & 0.049\\
7 & $d_{\text{WPSP}}(m,c)$&  $0.444\pm0.32$     & $0.899\pm0.72$  & 0.034 \\
\hline
\multicolumn{2}{c|}{Distances fusion : avg(4,5,6,7)}
                    & $0.500\pm0.138$ & $0.664\pm0.115$ & 0.001 \\
\hline
\end{tabular}
\end{table}

These results are important because, although non-fallers in average completed the TUGs faster than fallers (see Table~\ref{tab:sampledescription}), there is no significance between the means of the groups. We believe that the sequence of activities in a TUG test (standing, walking, turning, walking, sitting) carries a richer composition of frequencies, which we believe was captured by PSE and WPSP and the distance-based features. 

The PSE measure the complexity of the gait signals (via Entropy), which is higher for non-fallers. The WPSP is related to the harmonic components of the gait. It is interesting to note that the first harmonic (the fundamental frequency and amplitude that represents the gait) did not differ among the groups, which is probably due to the fact that all participants are active and non fragile. However, they differ when looking at the second and third harmonics, that are related to higher frequencies. Again, non-fallers have second and third harmonics in larger amplitude and frequencies, when compared to fallers.

Regarding the distance-based features, those measure how the features differ among different TUG tests, by computing the distance between values extracted from the signals.  We observed that the values showed larger differences between signals for non-fallers when compared to fallers. This indicates hat the different TUGs altered more the gait pattern of non-fallers during the tests.

As occurred with the TUG test seconds in our study, in previous works, speed gait only was also shown to be insufficient to identify fallers~\cite{Ponti2015learning}, which might indicate the importance of analysing different activities. According to our results, by pre-processing the TUG signals and extracting both individual frequencies features (PSE, WPSP) and other comparative features, it is possible to obtain a set of features with significant difference between the fallers and non-fallers.

\begin{figure*}[hbt]
\centering
    \begin{tabular}{c}
    \includegraphics[width=0.9\linewidth]{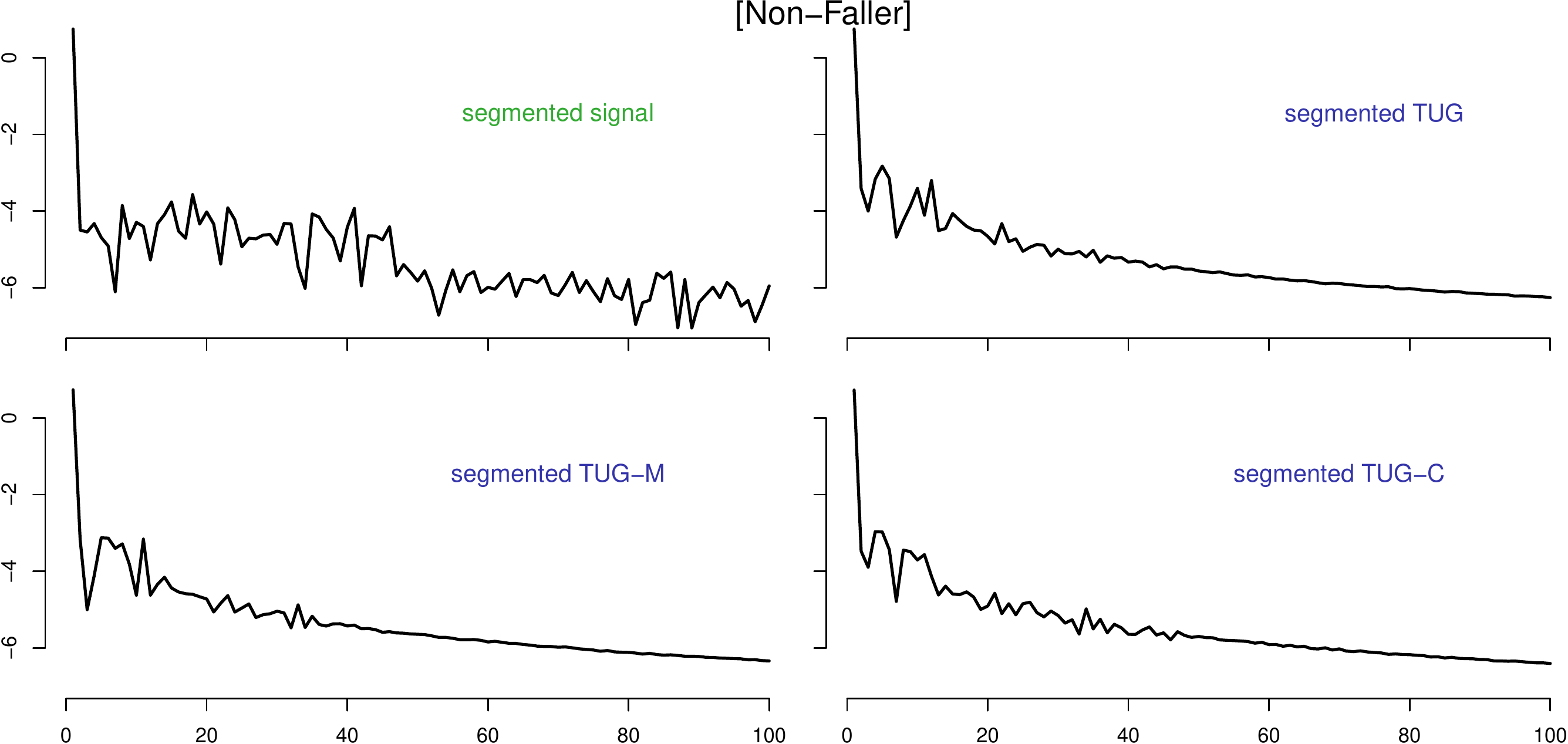} \\[12pt]
    \includegraphics[width=0.9\linewidth]{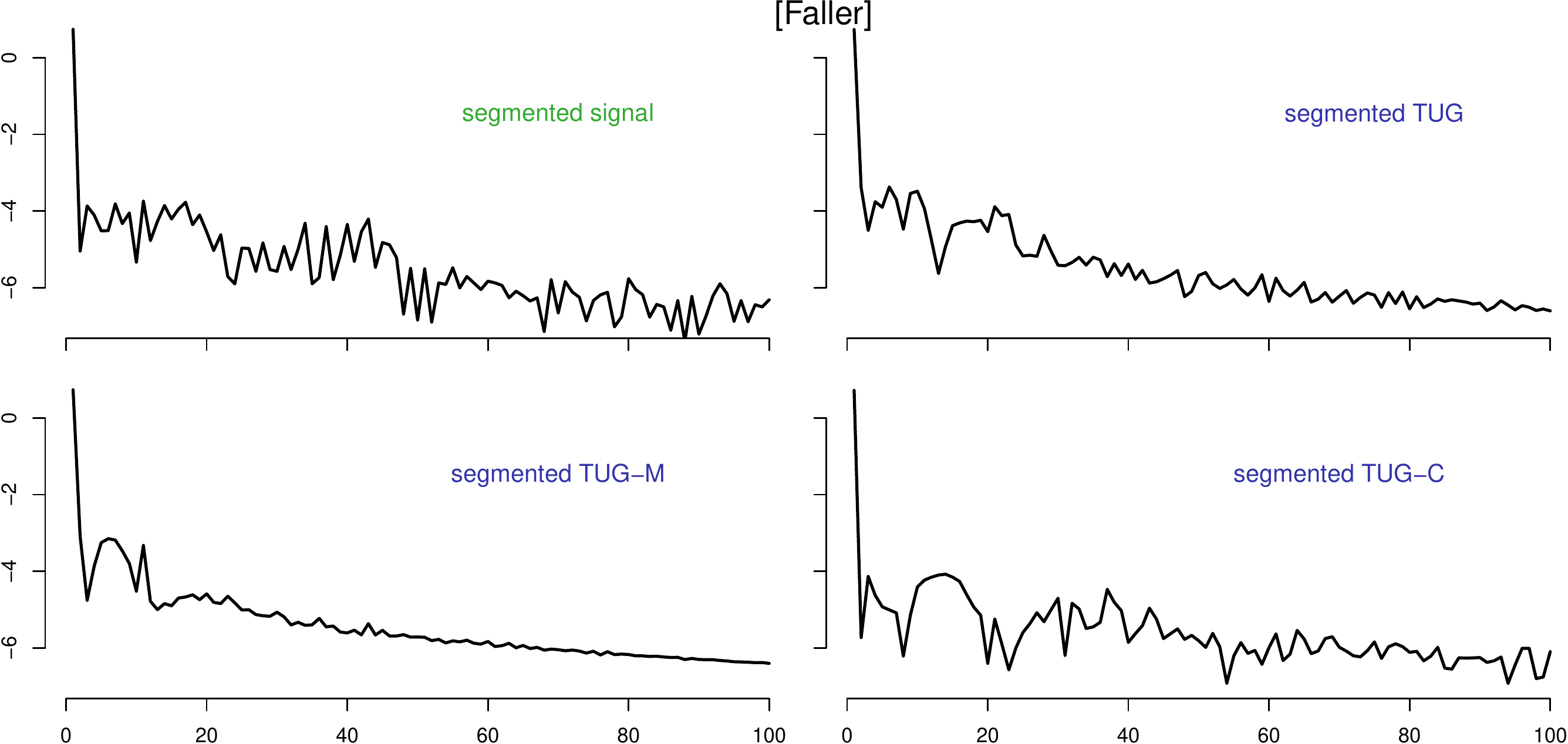}
    \end{tabular}
\caption{Power spectra examples of non-faller and faller with frequencies up to 100Hz for different signals used in this study to extract features: the three TUGs (segmented signal); only the regular TUG (segmented TUG); only the dual task manual TUG (segmented TUG-M); and only the dual task cognitive TUG (segmented TUG-C). The amplitudes are in a logarithmic scale.}
\label{fig:spectra}
\end{figure*}

In Fig~\ref{fig:spectra} we show an example of power spectra of signals acquired from a non-faller and a faller participant. The power spectra shows the profile of frequencies present in the signal. A careful inspection comparing those profiles reveal the differences between Fallers and Non-Fallers. In particular the TUG and TUG-C power spectra are more discriminative between the groups. In addition, it can be seen in this example how the distribution of frequencies is more diverse among the signals in the case of the faller, while in the non-faller case the frequency features for TUG, TUG-C, TUG-M are more similar. This corroborates our results since the feature extraction step is based upon signal frequencies.

\subsection*{ROC curve analysis}
\label{ss.rocanalysis}

\begin{figure}
\centering
    \setlength\tabcolsep{2pt}
    \begin{tabular}{cc}
    \includegraphics[width=.46\linewidth]{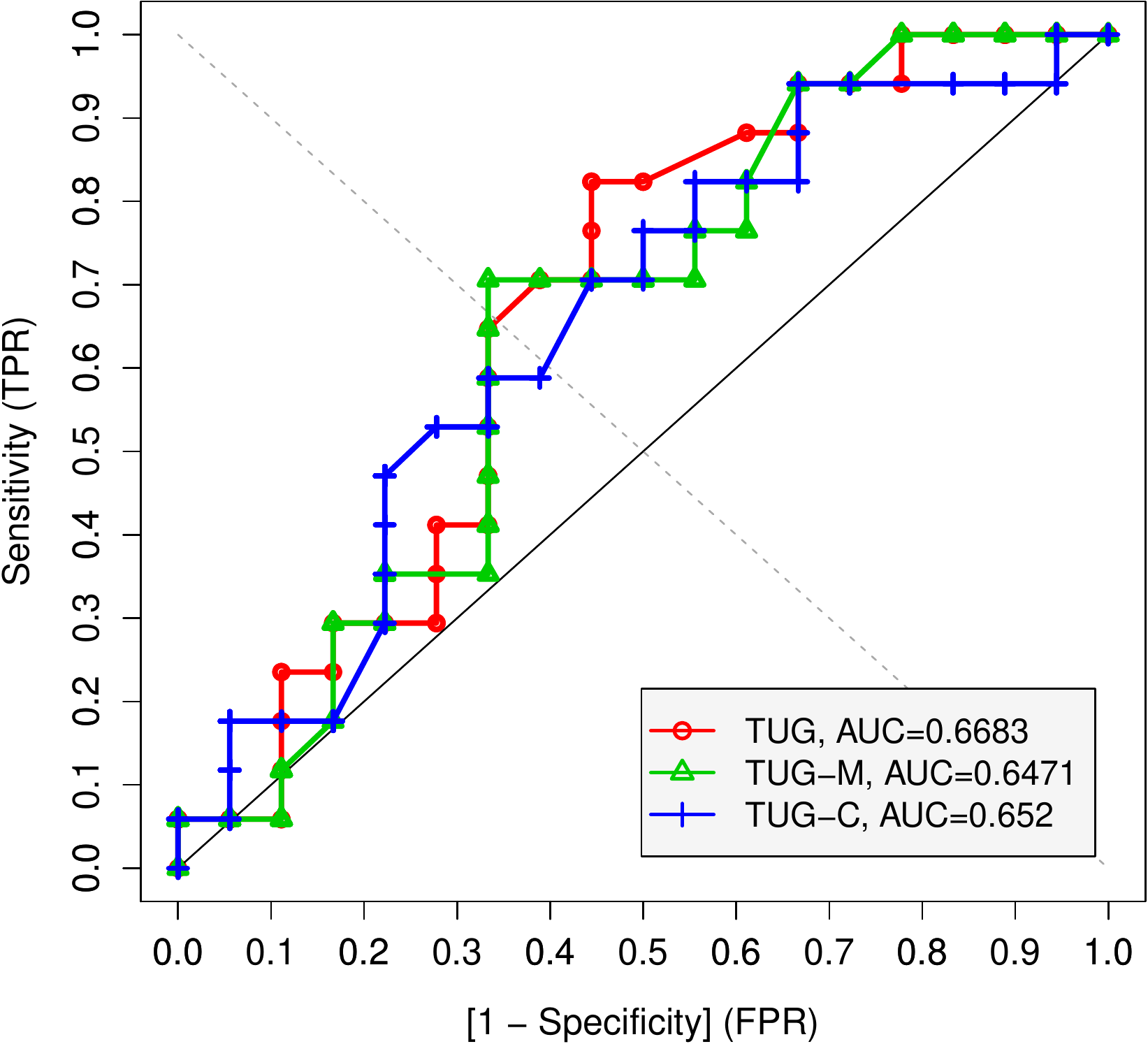} &
    \includegraphics[width=.46\linewidth]{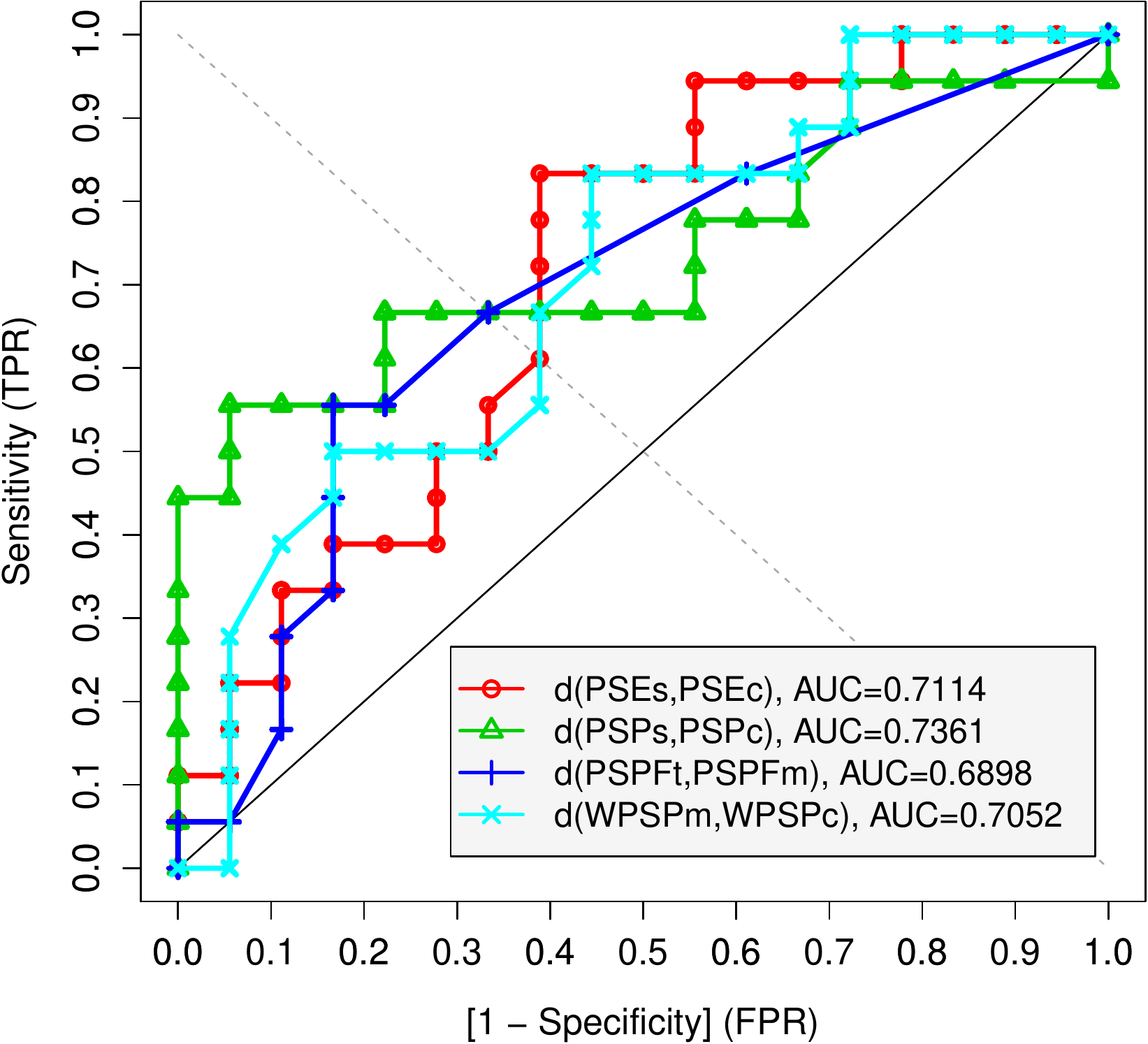} \\
    (a) TUGs (seconds) & (b) Distance Features\\[12pt]
    \includegraphics[width=.46\linewidth]{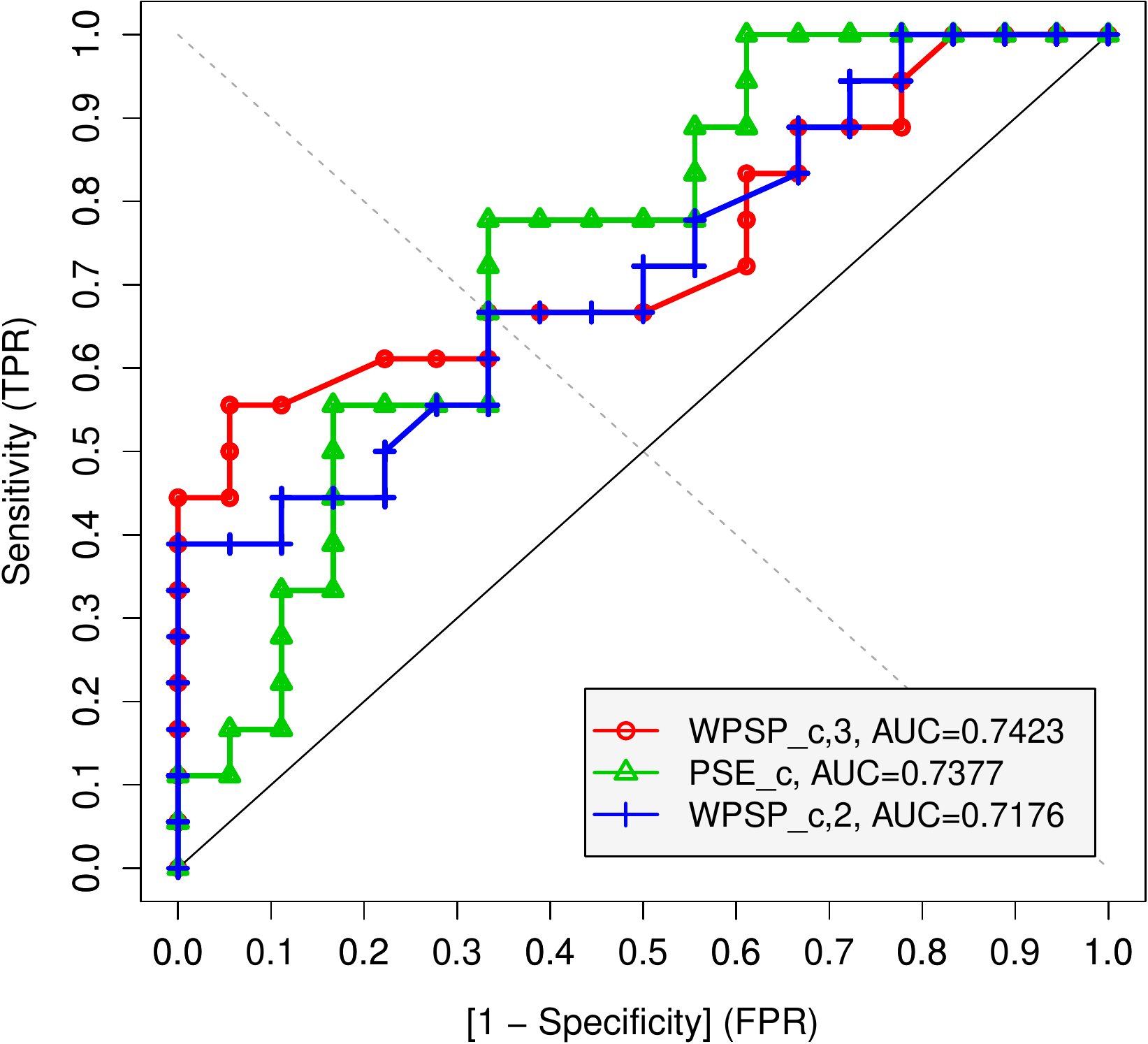} &
    \includegraphics[width=.46\linewidth]{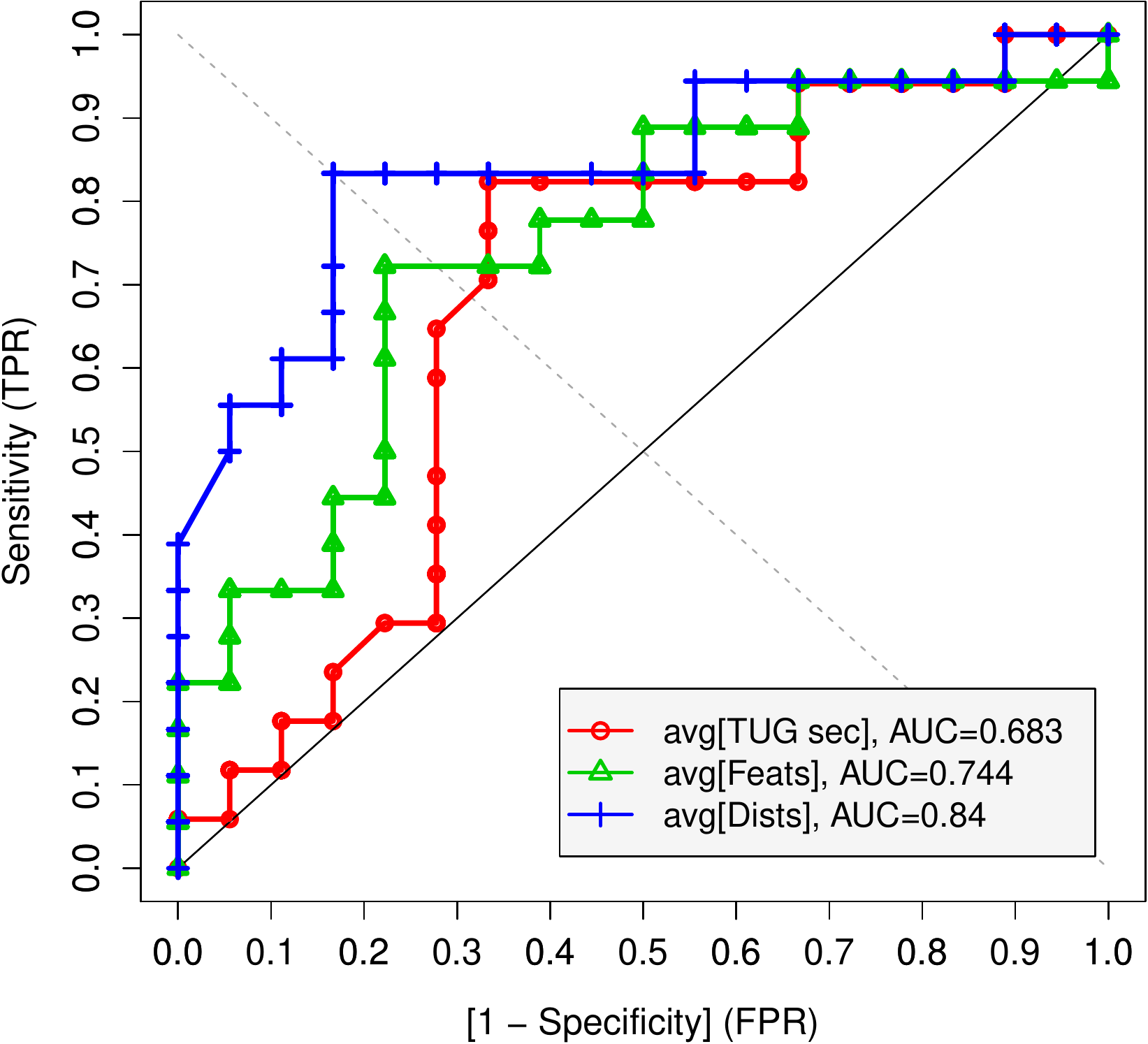} \\
    (c) Frequency Features & (c) Fusion methods \\
    \end{tabular}
\caption{ROC curves comparing the TUG seconds, Frequency-based, Distance-based and Fusion methods.}
\label{fig:ROCcurves}
\end{figure}

In order to compare the diagnosis capability of features found to be significant with the regular functional tests based on TUG and dual-task TUGs, we performed ROC analysis showing the curves in Figure~\ref{fig:ROCcurves}. A ROC curve is a compact representation of the result, showing how the use of different threshold (cut-off) values of a given variable impact on the Sensitivity and Specificity. We desire a curve that is higher, approaching values of 1.0 for both Sensitivity and Specificity. This curve will also have a higher value of Area Under the Curve (AUC). The ROC plot was chosen since it is considered the most stabilised analysis of assessing and using diagnostic tools, unifying the process to compare the tests in clinical medicine and health sciences~\cite{zweig1993receiver}. More details about ROC analysis and its derived measures can be found in~\cite{Fawcett2006}.

\begin{figure}
\centering
\setlength\tabcolsep{2pt}
    \begin{tabular}{cc}
    \includegraphics[width=.42\linewidth]{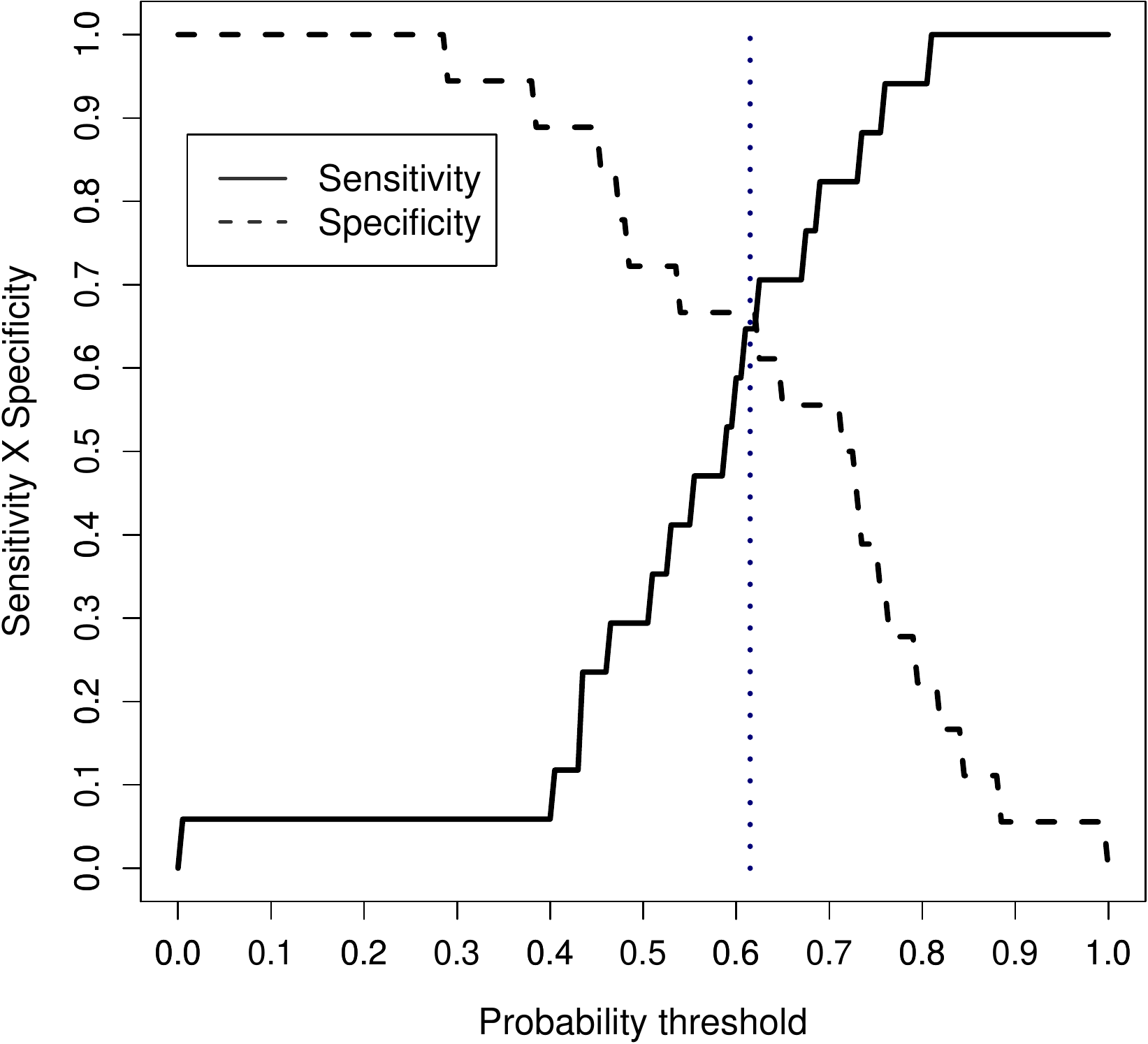} &
    \includegraphics[width=.42\linewidth]{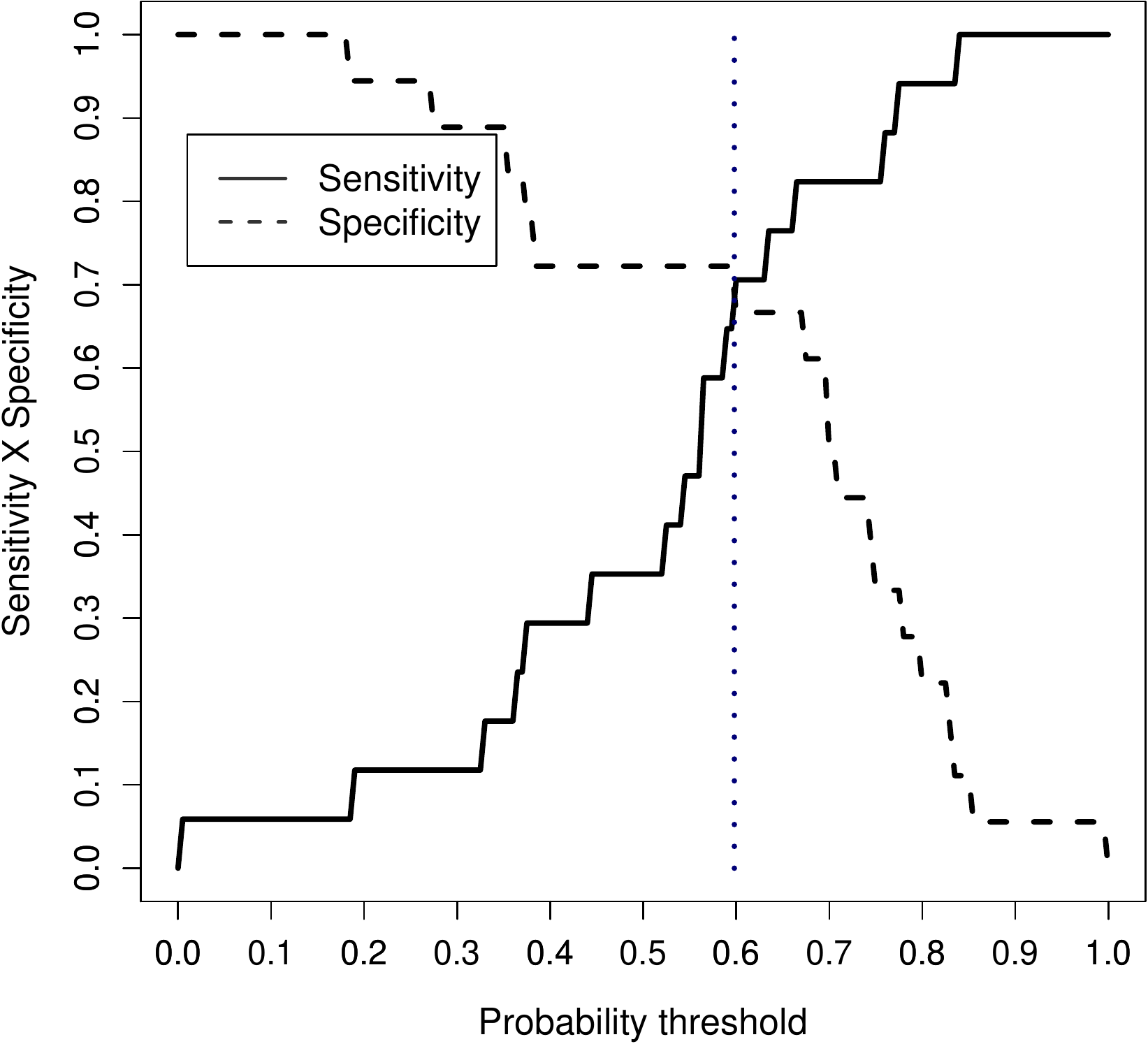} \\
    (a) TUG  & (b) avg[TUG]\\[12pt]

    \includegraphics[width=.42\linewidth]{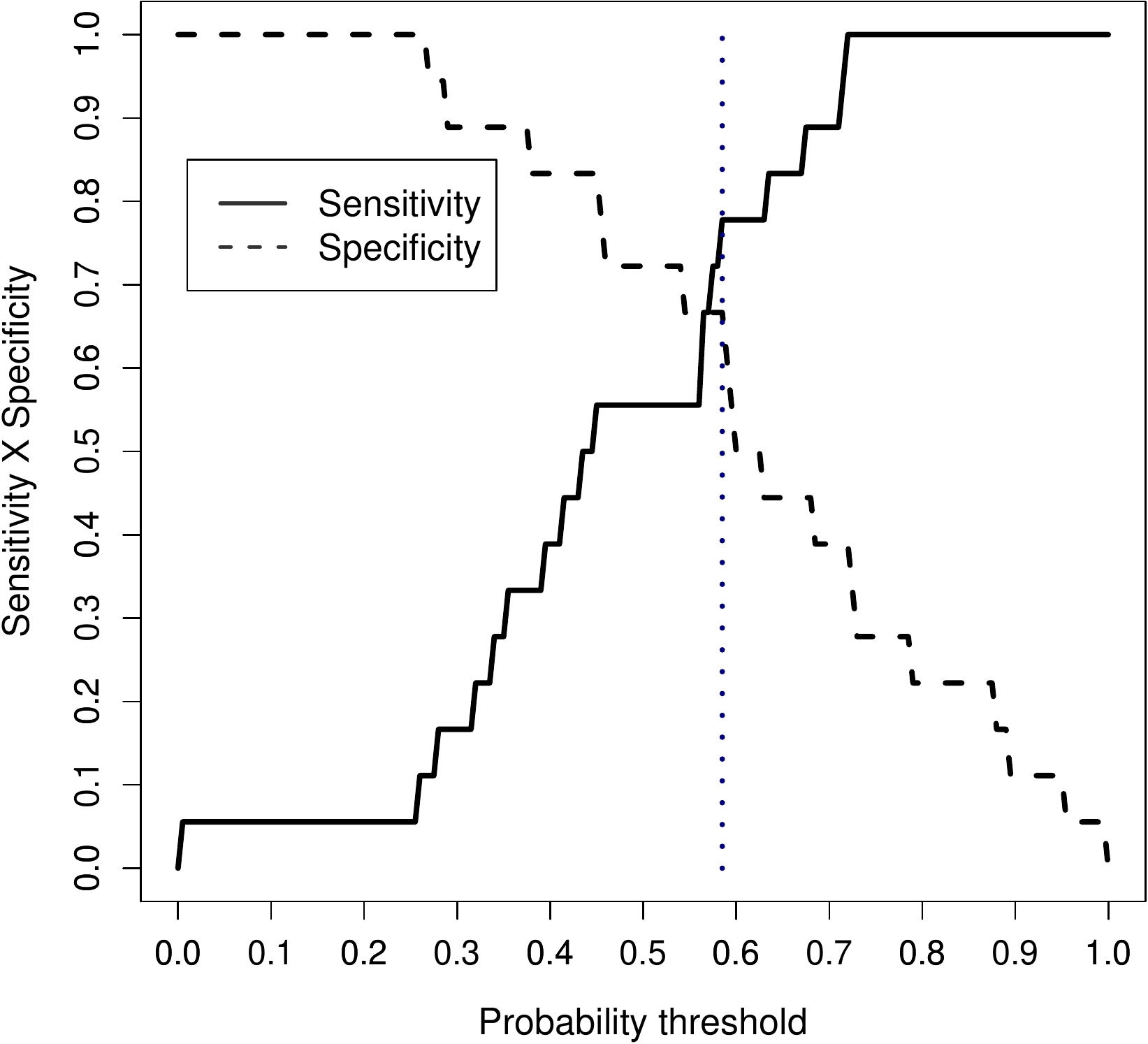} &
    \includegraphics[width=.42\linewidth]{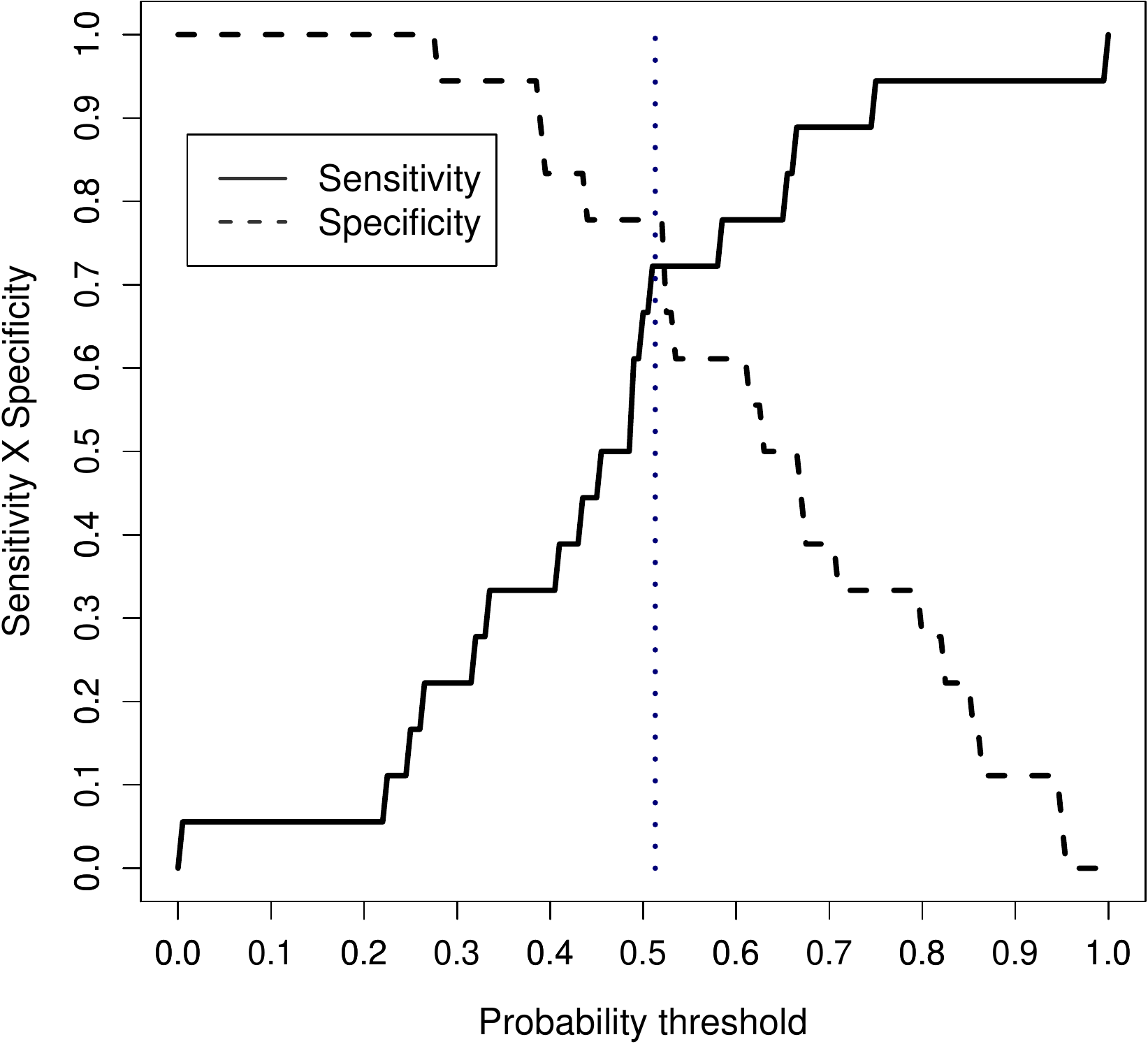} \\
    (c) PSE-c & avg[Feats] \\[12pt]

    \includegraphics[width=.42\linewidth]{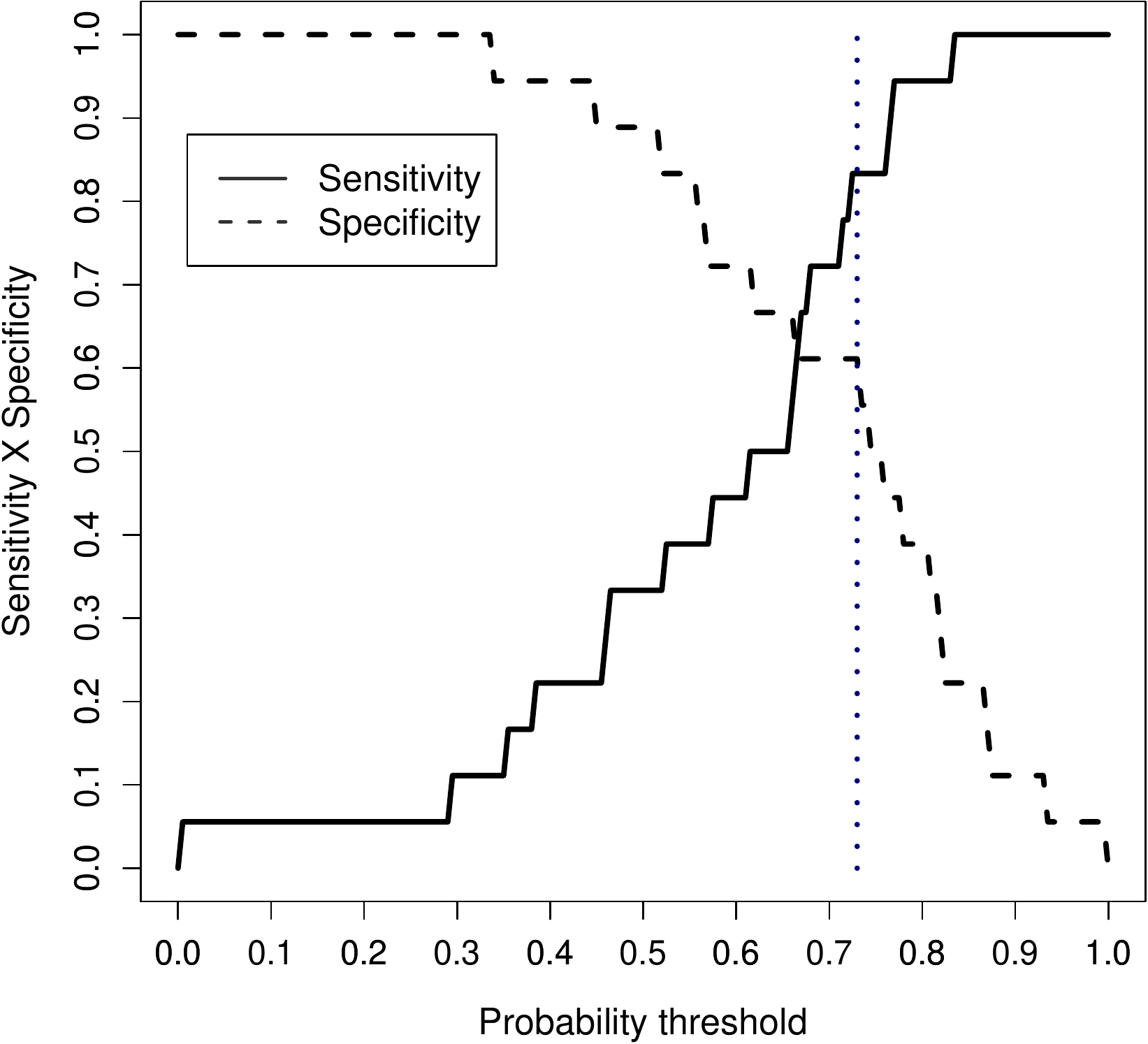} &
    \includegraphics[width=.42\linewidth]{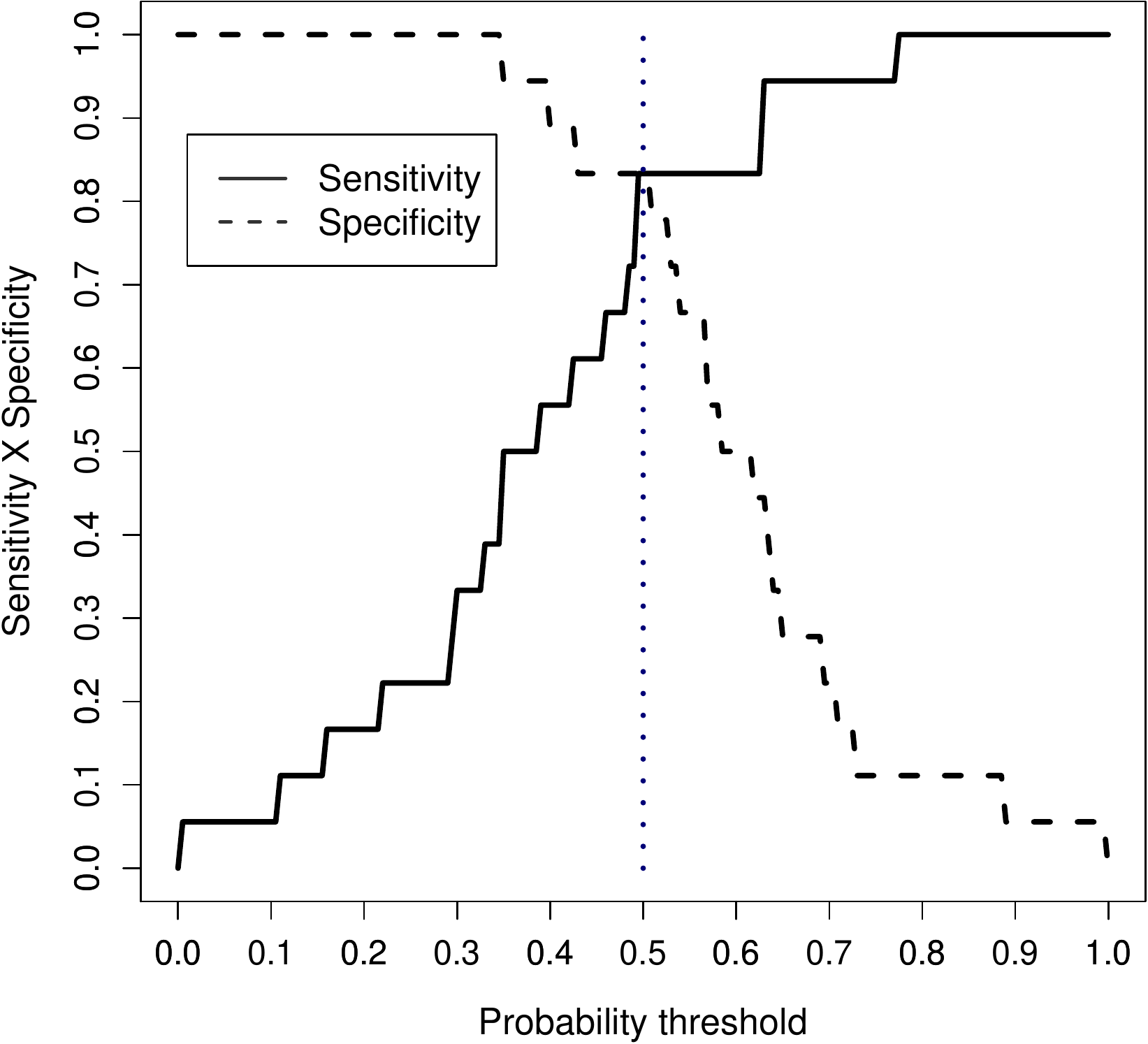} \\
    (c) d(PSE-s,PSE-c)  & (d) avg[Dists] \\[12pt]
    \end{tabular}
\caption{Sensitivity vs Sensitivity Analysis of the most relevant variables.}
\label{fig:SSanalysis}
\end{figure}

The values of AUC, Sensitivity (TPR), Specificity (1-FPR), and f1-Score for each individual and combined variable are shown in Table~\ref{tab:ROCanalysistable}. The values for Sensitivity, Specificity, and f1-Score were computed using the optimal probability cut-off threshold of a Sensitivity vs Specificity analysis as shown in Figure~\ref{fig:SSanalysis}.

The ROC analysis corroborates the findings of the previous section: the extracted Frequency and the Distance-based features have higher AUC and f1-Scores, achieving higher values for Sensitivity and Specificity. In particular, the fusion of features showed the best results. The average of the TUGs achieved AUC$= 0.68$, Sensitivity$=$Specificity$=0.70$, which is similar to those obtained in~\cite{Alexandre2012}.

\begin{table}[hpbt]
\caption{ROC analysis comparing TUG seconds with the accelerometer features}
\label{tab:ROCanalysistable}
\centering
\begin{tabular}{c | r r r r | r r}
 \hline
              & AUC     & TPR$^1$ &  1-FPR$^1$ & f1-Score & pr. cut-off & val. cut-off\\
 \hline
 \hline
TUG           &  0.668  & 0.64   & 0.64   & 0.64 & 0.61 & 8.73s\\
TUG-M         &  0.647  & 0.64   & 0.64   & 0.64 & 0.69 & 8.90s\\
TUG-C         &  0.652  & 0.58   & 0.67   & 0.58 & 0.65 & 11.31s\\
 \hline
\em TUGs avg       &  0.683  &\em0.70   &\em0.70   &\em0.70 & 0.60 & 10.17s\\
 \hline
\hline
PSE-$c$       &  0.737  &\em0.78 & 0.67   &\em0.74 & 0.58 & 11.26\\
WPSP-$c$,2    &  0.742  & 0.67   & 0.67   & 0.67 & 0.31 & 1.4508\\
WPSP-$c$,3    &  0.717  & 0.67   & 0.67   & 0.67 & 0.32 & 1.6554\\
\hline
\em Feats avg     &  0.744  &\em0.73   &\em0.78   &\em0.74 & 0.51 & 4.706\\
\hline
\hline
$d_{\text{PSE}}(s,c)$ & 0.711  &\em 0.83 & 0.61 &\em 0.75 & 0.73& 2.970\\
$d_{\text{PSP}}(s,c)$ & 0.736  & 0.67    &\em 0.78 &\em 0.71 & 0.66& 0.237\\
$d_{\text{PSPF}}(t,m)$& 0.690  & 0.67    & 0.67 & 0.67 & 0.92& 11\\
$d_{\text{WPSP}}(m,c)$& 0.705  & 0.67    & 0.61 & 0.65 & 0.18& 0.211\\
\hline
\em Dists avg     & 0.840 & \em 0.83 & \em 0.83  & \em 0.83 & 0.50 & 0.5786\\
\hline
\multicolumn{6}{l}{\scriptsize OBS: values of TPR and 1-FPR are relative to the cutoff point;} \\
\multicolumn{6}{l}{\scriptsize values highlighted in italic are those higher then 0.67} \\
 \hline
\end{tabular}
\end{table}

Although the individual features performed better then the TUG variables, it was the fusion of the variables that was able to reach the best results. The fusion of the three frequency features achieved AUC$=0.744$, Sensitivity$=0.73$ and Specificity$=0.78$ with f1-Score$=0.74$, while the fusion of the four distance-based features achieved AUC$=0.84$, Sensitivity$=$Specificity$=0.83$ for the probability cut-off point $0.5$ with a 95\% CI $0.62$-$0.91$. This result is due to the fact that these variables were complementary. This effect can be seen in the ROC curves. For example, when inspecting the Distance Features in Figure~\ref{fig:ROCcurves} (b), while the d(PSEs,PSEc) --- red curve --- shows to privilege specificity, the d(PSPs,PSPc) --- green curve --- seems to privilege Sensitivity. Similar effects also occur with the other distance-based features and within the frequency features, and for this reason by combining the variables it is possible to better discriminate the groups.

There are many relevant differences when comparing our study with related work in the literature. Salarian et al. used five to seven inertial sensors in contrast with our single sensor input \cite{salarian2010itug}; Zampieri et al. focused on participants with Parkinson's Disease\cite{zampieri2011assessing}, while our sample is composed of healthy elderly; Weiss et al. investigated step-to-step consistency during 3 days~\cite{Weiss2013}, instead of a TUG test as used in our paper. Sejd\'ic et al.~\cite{Sejdic2014} found no difference between the faller and non-faller groups using gait frequency features extracted from the accelerometer, whereas in our study, the differences between TUGs and the cognitive component during gait and other sub-components of the TUG tests were more sensitive and specific to identify fallers.
Weiss et al.~\cite{Weiss2011} conducted a study in which accelerometer signals were analysed from three consecutive days using an accelerometer and performing activities of daily living without supervision instead of using the TUG dual task. In their study, as in ours, the frequency amplitudes from acceleration signals showed statistical difference among the elderly fallers and non-fallers, corroborating the potential of frequency features as a viable feature to identify fallers. 

Muhaidat et al.~\cite{muhaidat2013exploring} observed that elderly fallers were slower when compared to non-fallers during walking under a triple task tests with a cognitive component. In this case, they concluded that fall history are correlated with changes in both gait features and balance. Although the number of steps showed no difference between groups in our results using the single and dual task TUG, the cognitive task was sufficient to show differences capable of discriminating fallers from non-fallers. Finally, in van Schooten et al. several features based on quantity and quality of gait were combined with input from questionnaires, grip strength and trail making test achieving a AUC of $0.82$ in a prospective study\cite{vanSchooten2015}, whereas we focused only on input from accelerometer data achieving a similar value of AUC=$0.84$, which indicates that the TUG cognitive may be a better test than using only free gait for identify fallers.

\section{Conclusions}
\label{s.conclusion}

In this paper, we address the problem of faller identification in a sample of healthy older individuals that cannot be classified into fallers and non-fallers by using only functional mobility tests. While the features extracted from the whole signal are not discriminative (from both the statistical test and the ROC analysis perspectives), features extracted from dual-task cognitive TUG, as well as the distance-based features shows statistical difference. By combining the features via average fusion, we were able to increase the discriminative power, reaching AUC = 0.84 and values for Sensitivity and Specificity of 0.83.

With respect to previous work in the literature, we highlight the use of a single accelerometer instead of multiple sensors, active and healthy participants, frequency-based features instead of step-based variables and the inclusion of fusion of variables based on the different single and dual task tests.

A limitation of the study is the sample size (36 participants). In addition, due to the intra-class variability of the data, it is not possible to generalise this result for a broader population. However, the results does indicate that features based on dual-task TUGs are able to better discriminate faller and non-fallers, even in a scenario when all standard tests and measures were insufficient to show significant differences. Furthermore, both distance-based features and fusion shown to be interesting methods to improve the results. Finally, the acquired dataset is available to be used in future investigations.

Future studies could use our methods within a free gait data collection study by first detecting sitting and standing activities, which are present in TUG tests, and processing those signals in order to extract the features. Also investigating orientation-based features, by considering each axis separately is a matter of future work.

\section*{Acknowledgments} 
We would like to thank all the participants of this study as also the University of the Third Age (Universidade Aberta da Terceira Idade de S\~ao Carlos - FESC). This work is partially funded by FAPESP grants \#2015/13504-2, \#2016/10982-3, \#15/09715-8, \#14/21858-6; and CNPq grant \#2013-2/458793. The authors greatly appreaciate their support.


\end{document}